\documentclass[12pt,preprint]{aastex}
\singlespace
\slugcomment{Accepted for the publication in the {\it Astrophysical Journal}}
\begin{document}
\def\eg{{\it e.g.}}
\newcommand{\tnm}[1]{\tablenotemark{#1}}
\newbox\grsign 
\setbox\grsign=\hbox{$>$}
\newdimen\grdimen 
\grdimen=\ht\grsign
\newbox\simlessbox 
\newbox\simgreatbox
\setbox\simgreatbox=\hbox{\raise.5ex\hbox{$>$}\llap
     {\lower.5ex\hbox{$\sim$}}}\ht1=\grdimen\dp1=0pt
\setbox\simlessbox=\hbox{\raise.5ex\hbox{$<$}\llap
     {\lower.5ex\hbox{$\sim$}}}\ht2=\grdimen\dp2=0pt
\def\simgreat{\mathrel{\copy\simgreatbox}}
\def\simless{\mathrel{\copy\simlessbox}}

\title{The Recent Cluster Formation Histories of 
NGC~5253 and NGC~3077: Environmental Impact on Star Formation
\altaffilmark{1}}

\author{Jason Harris}
\affil{Space Telescope Science Institute}
\affil{3700 San Martin Dr., Baltimore, MD, 21218}
\affil{E-Mail: jharris@stsci.edu}
\author{Daniela Calzetti}
\affil{Space Telescope Science Institute}
\affil{3700 San Martin Dr., Baltimore, MD, 21218}
\affil{E-Mail: calzetti@stsci.edu}
\author{John S. Gallagher III}
\affil{Dept. of Astronomy, University of Wisconsin-Madison}
\affil{475 North Charter Street, Madison, WI 53706}
\affil{E-Mail: jsg@astro.wisc.edu}
\author{Denise A. Smith\altaffilmark{2}}
\affil{Space Telescope Science Institute}
\affil{3700 San Martin Dr., Baltimore, MD, 21218}
\affil{E-Mail: dsmith@stsci.edu}
\author{Christopher J. Conselice}
\affil{California Institute of Technology}
\affil{MC 105-24, Pasadena, CA 91125 }
\affil{E-Mail: cc@astro.caltech.edu}

\altaffiltext{1}{Based on observations obtained with the NASA/ESA 
{\it Hubble Space Telescope} at the Space Telescope Science 
Institute, which is operated by the Association of Universities for 
Research in Astronomy, Inc., under NASA contract NAS5-26555.}

\altaffiltext{2}{Computer Sciences Corporation}

\begin{abstract}
We present multicolor photometry of bright star cluster candidates in
the nearby starburst galaxies NGC~3077 and NGC~5253, observed with 
the {\it Hubble Space Telescope} Wide Field and Planetary Camera~2 in 
both broadband (F300W, F547M and F814W) and narrowband (F487N and 
F656N) filters.  By comparing the photometry with theoretical 
population synthesis models, we estimate the age and mass of each 
star cluster, which provides constraints on the recent star formation 
histories of the host galaxies.  We compare the star cluster 
populations in these dwarf starburst galaxies to those of the nuclear 
starburst in the barred spiral M~83, and discuss implications for our 
understanding of the nature and evolution of starburst events.
\end{abstract}

\keywords{galaxies: starburst --- galaxies: star clusters ---
galaxies: individual: NGC 5253 --- galaxies: individual: NGC 3077 }

\section{Introduction}\label{sec:intro}

Throughout the Universe's history, a significant fraction of star
formation has apparently occurred in high-intensity bursts.  These
rapid star-formation events play important roles in the ionization,
chemical enrichment, and overall evolution of their host galaxy.
Such starburst events are often seen in interacting galaxies,
and it seems reasonable to infer a causal connection between galaxy
interactions and starburst activity.  However, interactions are 
probably not the only causal agent for starburst activity; internal 
dynamics can also play an important role (\eg, M~83).  
Combined with the dynamic merger histories predicted by the standard 
hierarchical collapse model of galaxy formation, it is likely that 
much of a typical galaxy's stellar content was formed during 
starburst events.  Despite this ubiquity, we currently have only a 
rudimentary understanding of the cause, nature and evolution of 
starburst events, and in particular, of the physical processes which 
connect the external environment to the internal star formation rate.

Nearby starburst galaxies provide important laboratories for
investigating the processes involved in galactic-scale star-formation
events in detail.  Local starburst galaxies appear to be similar to 
the active galaxies seen at high-redshift ($z\simgreat2.5$) in 
UV colors \citep{meu97, mhc99, as00} and spectral morphology 
\citep{ste96, low97, pet00}, although the high-redshift systems 
typically have star-formation rates that are several times higher 
than local starbursts.  Because of their similarity, we can 
reasonably expect that detailed photometric and spectroscopic studies 
of local starbursts will lead to a general understanding of the 
starburst phenomenon at all redshifts.  In particular, we wish to 
address the following questions regarding the physical processes 
governing the starburst phenomenon: 
(1) Are starburst properties determined by the global properties of
the host galaxies?  
(2) How long do starbursts last?
(3) What are the mechanisms that cause and sustain starbursts?
(4) Does star-formation activity propagate across galaxies?  If so,
what is the propagation mechanism?
(5) On what timescale do star clusters dissolve in starbursts?
(6) Is the field stellar population composed primarily of evaporated
cluster stars, or is there a significant population that is native to
the diffuse field?

To investigate these questions, we are undertaking a multi-band
photometric survey of nearby starburst galaxies using the Hubble 
Space Telescope ({\it HST}) Wide Field and Planetary Camera 2 ({\it 
WFPC2}).  The program's goal is to survey the stellar populations and 
gas content of nearby starburst galaxies spanning a range of internal 
properties (mass, metallicity, internal dynamics) and external 
environments, in order to understand better what role these factors 
play in determining the nature and evolution of the galaxies' 
starbursts.  The focus of the present work is to constrain the recent 
star-cluster formation histories of two nearby dwarf galaxies 
(NGC~3077 and NGC~5253) which have similar mass ($M\sim10^9M_\odot$) 
and luminosity ($L_B\sim10^9L_{B\odot}$), but large contrasts in 
metallicity and external environment.  Until recently, it was thought 
that NGC~3077 and NGC~5253 were both interacting satellites of large 
spiral galaxies.  NGC~3077's membership in the M~81 group is not in 
doubt, but recent Cepheid distance measurements place NGC~5253 about 
600~kpc from its nearest neighbor, M~83 \citep{kar02b, sch94}, 
making it unlikely that M~83 has influenced NGC~5253's recent 
history.

We also compare their recent star-cluster formation histories to that 
of the large spiral galaxy M~83 (NGC~5236), which was studied in 
Paper~1 of this series \citep{har01}.  The global properties of 
NGC~3077, NGC~5253 and M~83 are compared in Table~\ref{tab:galprop}.  

The paper is organized as follows.  We present the observations and
data reduction in Section~\ref{sec:data}.  In 
Section~\ref{sec:analysis}, we present the photometric catalogs, the 
extinction corrections, the star/cluster separation analysis, and our 
estimates of the star cluster ages and masses, based on both 
broad-band photometry and the equivalent width of the $H\alpha$ 
emission line.  We interpret and discuss the results in 
Section~\ref{sec:discuss} and summarize the work in 
Section~\ref{sec:summary}.

\section{Observations and Data Reduction}\label{sec:data}

The nearby dwarf starburst galaxies NGC~3077 and 
NGC~5253 were observed with the {\it HST} {\it WFPC2} instrument, 
in the broad-band filters F300W, F547M, and F814W (see 
Figure~\ref{fig:color}), and in the narrow-band filters F656N and 
F487N on three visits between 1996 and 2001 (see 
Table~\ref{tab:observations} for exposure information).  
Images were also taken in the narrow-band filters F502N and F673N; 
these data are discussed in a companion paper \citep{cal03}.  In each 
exposure, the starburst region was centered on the WF3 chip.  This 
provided a field of view wide enough to cover the optical starburst 
region in each galaxy, at the expense of the poorer angular 
resolution of the WF chip.

The data were reduced by the STScI calibration pipeline, which 
includes flagging of bad pixels, A/D conversion, bias and dark 
current subtraction, and flatfielding. Hot pixels were removed or 
flagged using the STSDAS routine WARMPIX, which uses hot pixel 
information from dark frames obtained around the time of the science 
observations to perform the correction. 

In each of the filters, two or three separate exposures were obtained
to aid in cosmic ray rejection.  Cosmic ray rejection and co-addition 
were performed using the STSDAS routine CRREJ \citep{wil96}, with a 
rejection threshold of 4~$\sigma$ for the cosmic rays and 
2.4~$\sigma$ for the adjacent pixels.  NGC~3077's CR-split exposures 
in F547M and F814W have orientations which differ by approximately 
0.5\arcdeg (see Table~\ref{tab:observations}), and there is also a 
non-integer pixel translation offset between these exposures.  These 
registration offsets degrade the effective PSF of the F547M and F814W 
images after registering them for CR cleaning.  Visual inspection of 
the radial profiles of star-like objects indicates that the FWHM is 
typically enlarged by 40--50\% in the CR-cleaned images, compared to 
the original CR-split exposures.  While this does not have a large 
effect on our photometry, our star/cluster determination analysis 
must be performed on the CR-split images, where the full angular 
resolution of WFPC2 has not been compromised (see 
Section~\ref{sec:starclust}).

The absolute photometric calibration of the images is obtained from 
the zero-points listed in HST Data Handbook \citep{hst02}, and have 
about 2--5\% accuracy in the medium and broad-band filters 
\citep{cas97}.  The effect of contaminant buildup onto the WFPC2 
window is negligible at optical wavelengths, but not for F300W.  
The F300W images were taken 7 and 13 days after the previous 
decontamination procedure (for NGC~3077 and NGC~5253, respectively), 
so we applied contamination flux corrections of 1.6\% and 3\%.
The well-known Charge-Transfer Efficiency (CTE) problem of WFPC2 
\citep{ste98, whc99, dol02} presents some challenges for our 
analysis.  Most of the analysis of the CTE problem to date has 
focused on point sources; however, recently \cite{rie00} studied the
CTE effect on extended objects, and found that the reduction of flux 
was greatly mitigated compared to point sources.  The primary effect
on extended sources is to make the profile shape asymmetric in the 
direction of the readout.  Because our objects are slightly resolved,
we believe the standard CTE formulae derived for stars may be 
inappropriate in this case.  In addition, our broad-band images have 
an elevated background level, which reduces the CTE effect, because 
charge-traps are already filled by the background flux.  Even
in the narrow-band images, the background is highly variable, and our 
sources are preferentially found around the image center, which makes 
the ramp correction less significant.  In light of these
uncertainties, we simply apply a 2\% correction to the total counts
in our narrow-band images, approximatel equal to the stellar CTE 
correction for the center of a WFPC2 chip.  

The construction of line-emission $H\alpha$ and $H\beta$ images from 
the narrow-band F656N and F487N images is discussed in detail in 
\cite{cal03}.  Briefly, we subtract off the underlying stellar 
continuum emission and remove contaminating [\ion{N}{2}] flux from 
the $H\alpha$ images.  We make an additional correction for 
underlying stellar absorption in the $H\beta$ images (stellar 
absorption is considered negligible for the $H\alpha$ images).

\section{Photometry and Analysis}\label{sec:analysis}

In \citet{har01}, we presented a method for determining the 
photometry of star clusters from WFPC2 images of nearby galaxies.  
Our method involves convolving each image with a small Gaussian 
kernel ($\sigma=1$--2~pixels), and determining a best-fit PSF-like 
cluster profile model for each image.  Profile-fitting photometry is
preferable to aperture photometry due to the crowded conditions of 
the frames, and to the highly variable backgrounds.  The Gaussian
convolution is necessary in order to homogenize the variable
cluster profiles to the point that they can be described well by a
single model.

In the present cases of NGC~3077 and NGC~5253, we found that the 
above procedure was insufficient for obtaining accurate cluster 
photometry.  Many of the objects left a circular discontinuity in the 
model-subtracted residual image, indicating a poor fit by the profile 
model.  Also, the objects showed a wide distribution of apparent 
aperture corrections, suggesting significant variation among the 
cluster profiles, despite our convolution with a Gaussian kernel.  We 
attempted to account for this variation by fitting two independent 
profile models: one optimized for broader clusters, the other for 
more compact clusters.  This improved the residual images 
significantly, but did not completely solve the variable-profile 
issues.

For these reasons, we have modified our cluster photometry method.
Since we cannot obtain a cluster profile model that fits all objects
well, we revert to aperture photometry, which does not require a uniform 
cluster profile shape.  However, we perform our aperture photometry 
on ``neighbor-cleaned'' images to avoid contaminating flux from 
nearby objects.  To construct neighbor-cleaned images, we first 
divide the list of objects into groups such that there are no pairs 
of objects with an angular separation less than 20~WF~pixels 
($\sim2\arcsec$) in any single group.  For each group, we then 
construct an image in which all objects except those in the current 
group have been fitted with the best (albeit imperfect) profile model 
and subtracted from the image.  The result is a series of images, 
each of which contains only isolated objects (see 
Figure~\ref{fig:subimage}).  We then measure aperture photometry 
for each object in the neighbor-cleaned image in which the object was 
left unsubtracted.  

We have cleaned the apertures of contaminating flux from neighboring 
objects, but the problems of variable background levels and 
non-uniform profile shapes remain, making it impossible to determine 
proper aperture corrections.  In our analysis of M~83, we were able 
to compute the mean aperture correction for artificial objects added 
to empty image regions, using the best-fit radial profile model.  
However, since the clusters in our current galaxies have non-uniform 
profile shapes (even after our attempt to homogenize them through 
convolution with a Gaussian kernel), we cannot derive a usable 
aperture correction from a model profile.  

Without aperture corrections, photometry suffers from systematic 
errors, since some unknown fraction of the object's total flux falls 
outside the photometric aperture.  The solution to this is to 
increase the photometric aperture size, thereby reducing the fraction 
of flux outside the aperture to a negligible level.  However, when 
the background is variable or otherwise poorly-determined, larger 
apertures add significant random errors to the photometry (which is 
why average aperture corrections derived from isolated objects are 
so useful).  We attempt a balance between the systematic errors 
imposed by small apertures and the random errors imposed by large 
apertures by adopting an intermediate aperture size of 5~pixels, 
which is approximately twice as large as a typical cluster's FWHM 
size.  

In Figure~\ref{fig:apcorr}, we show the 5-pixel aperture corrections 
for all detected objects in the three neighbor-cleaned broad-band 
images of NGC~3077.  The distribution of aperture corrections is 
very wide, confirming our assertion that variable cluster profile 
shapes and background levels make applying a standard background 
correction impossible.  However, we highlight objects which are 
located outside the dense, central regions of NGC~3077; the aperture
corrections of these more isolated objects are much more tightly 
distributed, especially among the brighter objects.  The mean 
aperture correction of the bright, isolated objects is 0.1--0.2~mag 
in all three filters, but the numbers of such objects are small, 
making a direct empirical determination difficult.  We simply adopt 
the 0.1~mag 5-pixel aperture correction appropriate for point sources, 
as a gross approximation of our mean aperture correction.  It is 
likely that there remains a systematic error in the photometry of 
order 0.1~mag associatied with the fact that the cluster profiles are
not point sources.

\subsection{Dust Extinction}\label{sec:dust}

As in our analysis of M~83, we rely on the ratio of $H\alpha$ to
$H\beta$ to estimate the extinction toward the clusters.  We 
construct an $H\alpha/H\beta$ ratio image after rebinning 
the emission-line images using a $3\times3$ boxcar kernel to remove 
small registration errors.  We compute a per-bin $E(B-V)$ estimate 
using the standard formula for dust in starburst galaxies 
\citep{cal97}.  We compare the distributions of inferred $E(B-V)$ 
values in NGC~3077 and NGC~5253 in Figure~\ref{fig:ebvhist}.  The 
average extinction in NGC~3077 is about twice that in NGC~5253.  In 
Figure~\ref{fig:ebvmap}, we present the $E(B-V)$ maps for NGC~3077 
and NGC~5253.  The maps illustrate the complex distribution of dust 
in these galaxies.  Note that we have imposed a 5$\sigma$ 
signal-to-noise cut on both $H\alpha$ and $H\beta$, which biases the 
maps against extremely high extinction values.  NGC~3077, in 
particular, contains some obviously highly extincted regions near its 
center that appear as zero-values in the $E(B-V)$ map.

We estimate the extinction toward each cluster by applying the same
two-component extinction model that we used for M~83:
\begin{eqnarray}
A_{300} = 3.07 \times (E(B-V)-E(B-V)_{MW}) + 5.59\times E(B-V)_{MW}\\
A_{547} = 1.79 \times (E(B-V)-E(B-V)_{MW}) + 3.10\times E(B-V)_{MW}\\
A_{814} = 1.14 \times (E(B-V)-E(B-V)_{MW}) + 1.79\times E(B-V)_{MW}
\end{eqnarray}
\citep{cal94, cal00, har01}.  $E(B-V)_{MW}$ is the foreground 
Galactic extinction estimate, equal to 0.07~mag for NGC~3077 and 
0.06~mag for NGC~5253 \citep{sfd98}.

\subsection{Star/Cluster Separation}\label{sec:starclust}

The biggest challenge in this analysis is in determining whether a 
given source is a star cluster, or an individual star.  We can easily 
make a luminosity cut at $M_V=-9$~mag (where $M_V$ is the 
extinction-corrected absolute magnitude in $V$), since individual 
stars are not expected to be more luminous than this \citep{whi99}.  
However, this cut alone will not remove contamination by foreground 
Galactic stars, and it will also remove real clusters, which may well 
be fainter than this luminosity.  

We therefore must rely on the high angular resolution provided by 
WFPC2 to 
provide shape information of the sources.  The WF PSF has a 
FWHM of about 0.13~arcsec, which corresponds to $\sim2$~pc at the 
characteristic distance of these galaxies (3.3--3.8~Mpc).  This is 
slightly smaller than the typical scale radius for young star 
clusters, so star clusters should be very slightly resolved in our 
images.

Since the clusters are only slightly larger than the instrumental 
PSF, we cannot use the Gaussian-convolved images which were used for 
determining the photometry; the star/cluster separation must be an 
independent analysis performed on the original, unblurred images.  We 
attempted a variety of methods to determine whether a given source 
was resolved or not, but eventually settled on a simple measurement 
of the objects' FWHM using the IRAF RADPROF tool, coupled with a 
visual examination of the objects' profiles using IRAF's IMEXAMINE 
tool.  

We note that in the case of NGC~3077, we could not use the cosmic-ray 
(CR) cleaned images for this analysis, because position-angle offsets 
among the CR-split images degrade the PSFs (see 
Table~\ref{tab:observations} and Section~\ref{sec:data} for details).
We instead performed FWHM measurements for NGC~3077 on the individual 
CR-split exposures.  In order to remove CR hits from our FWHM 
detection lists, we reject detections which are not present in each 
of the CR-split frames.  This provided us with two (or three) 
independent FWHM measurements for the real objects in each image.  We 
adopted the mean FWHM value for each object, unless it was 
determined that one of the values was affected by a coincident cosmic 
ray.

In Figure~\ref{fig:fwhm} we present the FWHM measurements in each 
band, as a function of extinction-corrected magnitude.  The 
magnitude corresponding to $M_V=-9.0$~mag is indicated; everything 
brighter than this is either a star cluster or a foreground Galactic 
star.  Note that objects brighter than the $M_V=-9.0$~mag cutoff are 
generally resolved, confirming that they are clusters.  Note also 
that NGC~3077 and NGC~5253 each contain only a handful of objects 
brighter than $M_V=-9.0$~mag; a marked contrast to M~83, in which we 
found 33 objects brighter than this limit.  

Upon inspecting the cluster profile shapes using the IRAF IMEXAMINE 
tool, we found that the FWHM values sometimes failed to reflect the
true cluster profile shape.  For example, an object with a FWHM 
indicating a PSF-like profile might have a radial profile that is 
clearly larger than a stellar profile.  We therefore base our final 
star/cluster classification on a manual comparison of the radial 
profile of each cluster candidate to that of a known star in the 
same image, using the measured FWHM values as a supplement and guide.  
Those objects whose profiles were unambiguously resolved 
in our visual inspection are flagged as clusters (indicated by solid 
circles in Figure~\ref{fig:fwhm}); we find 55 and 33 clusters in 
NGC~3077 and NGC~5253, respectively (see 
Figures~\ref{fig:n3077.clusterid} and \ref{fig:n5253.clusterid}).  
Objects whose profiles are indistinguishable from the stellar profiles 
are flagged as stars (indicated with star symbols in 
Figure~\ref{fig:fwhm}), and marginally-resolved or questionable objects 
are flagged as ambiguous (indicated by open circles in 
Figure~\ref{fig:fwhm}).  We did not attempt a detailed star/cluster 
classification for objects which were detected in only one of the three 
broad-band images; these partially-detected objects are represented as 
small points in Figure~\ref{fig:fwhm}.

\subsection{Determining Ages and Masses of the Star 
Clusters}\label{sec:agemass}

In Figure~\ref{fig:2cd}, we present the two-color 
diagrams for all objects in NGC~3077 and NGC~5253 which were detected 
in all three broad-band filters (65 objects in NGC~3077, and 117 in 
NGC~5253).  The Figure shows both the observed and 
extinction-corrected photometry of each object, connected by a 
straight line.  The objects which we determined to be resolved 
clusters are highlighted.  

We employ the Starburst99 population synthesis models \citep{lei99} 
to estimate the age and mass of each cluster.  We select 
instantaneous-burst Starburst99 models with an appropriate 
metallicity: Z=0.02 and Z=0.008 for NGC~3077 and NGC~5253 , 
respectively \citep{cal03, mar97}.  The models provide full 
spectral-energy distributions (SEDs); to obtain model {\it HST} 
photometry, we multiply the SED of each model point by the 
appropriate {\it HST} filter bandpass functions.  The model tracks 
are shown as the looping curve in each panel of Figure~\ref{fig:2cd}.  
Age increases along the model track from the upper left (bluer 
colors) to the lower right (redder colors).  The objects which we 
determined to be clusters follow the Starburst99 model track 
reasonably well.  Note that both NGC~3077 and NGC~5253 contain bright 
clusters whose photometry places them on the $\sim10$--20~Myr ``red 
loop'' portion of the model track, which is dominated by red 
supergiant stars.  The colors of these clusters have a large degree 
of intrinsic uncertainty, due to Poisson noise in the number of 
luminous red supergiants present in each cluster, and to uncertainties
in the evolutionary tracks themselves \citep[Starburst99 models are 
based on evolutionary tracks from the Geneva group; see\ ][for 
references]{lei99}.

The age estimate for each cluster is derived by matching the 
extinction-corrected photometry of each cluster to the point along 
the Starburst99 model with the most similar colors.  The effective 
``search area'' for each cluster traces an ellipse in the two-color 
diagram, extended along the reddening line, whose shape is determined 
by the $1\sigma$ extent of the photometric and reddening errors.  
This ``search area'' technique allows us to identify a plausible 
range of ages for each cluster, in addition to the best-fit age.  

Once we have estimated the age of a cluster, we then estimate its 
mass by assuming that the flux ratio of each cluster to its 
best-matched model point is equal to their mass ratio.  Each model 
point represents a stellar population with a mass of 
$1\times10^6$~M$_{\odot}$, but this assumes a Salpeter IMF over a 
limited mass range ($1.0 < M/M_\odot < 100.0$).  We apply a 
correction factor of 1.91 to obtain model cluster masses for the full 
range of stellar masses ($0.1 < M/M_\odot < 100.0$), using the 
cluster IMF presented by \cite{kro01}.  We obtain the plausible range 
of masses for each cluster by examining the flux ratio of the 
observed cluster to the model points at each end of the plausible age 
range.  The extinction-corrected photometry and age and mass 
estimates for each cluster candidate in NGC~3077 and NGC~5253 are 
presented in Tables~\ref{tab:n3077clusters} and 
\ref{tab:n5253clusters}, respectively.

\subsection{Age Constraints from $H\alpha$}\label{sec:halpha}

Because the flux-density of ionizing photons is a steeply decreasing 
function of stellar population age, the equivalent-width of the 
$H\alpha$ emission line ($EW(H\alpha)$) provides a reliable age 
indicator for stellar populations aged less than 10--20~Myr.  In 
order to construct $EW(H\alpha)$ maps of our WFPC2 fields, we use 
the same rebinned, [\ion{N}{2}]-cleaned, continuum-subtracted 
$H\alpha$ image that we constructed for our extinction analysis.  
The $H\alpha$ images for NGC~3077 and NGC~5253 are shown in 
Figure~\ref{fig:halpha}.  The positions of our cluster candidates 
are overplotted on the images.  

In order to assign an $EW(H\alpha)$ value to each cluster candidate, 
we perform simple aperture photometry at the position of each 
cluster, both in the $H\alpha$ image and in an $H\alpha$-continuum 
image.  The continuum image is constructed by interpolating between 
the F547M and F814W images, and renormalizing the interpolated image 
to the width of the F656N filter.  Since the $H\alpha$ and continuum 
images are binned 3x3, we first transform the clusters' pixel 
coordinates appropriately, and use a photometric aperture of 2 binned 
pixels.  The $EW(H\alpha)$ for each cluster is simply the ratio of 
the $H\alpha$ flux at the cluster position to that in the continuum 
image.  

We note that this procedure inherently assumes that all of 
the $H\alpha$ flux that is coincident with the cluster position is 
ionized by the cluster population itself.  It is troubling, in the 
context of this assumption, that there appears to be little 
correlation between cluster positions and peaks in the 
$H\alpha$-emitting gas.  In both galaxies (but especially in 
NGC~3077), some of the $H\alpha$ gas is distributed in large-scale 
filaments (see Figure~\ref{fig:halpha}), suggesting that the 
interstellar media of these galaxies is not necessarily opaque to 
ionizing photons.  In addition, supernovae and massive star winds in 
young clusters can evacuate their local region of interstellar gas, 
which could lead to an abnormally low $EW(H\alpha)$ measurement for 
a cluster of a given young age.  For these reasons, it is difficult 
to associate any lump of ionized gas with its ionizing source.
Nevertheless, young clusters do emit ionizing photons, so at some 
level there must be a correlation between local $H\alpha$ flux and 
the presence of an underlying young stellar population, despite 
these caveats.  We therefore retain our simple coincident-aperture 
$H\alpha$ measurement as a first-order indicator of the 
photoionization strength of each cluster.  

Tables \ref{tab:n3077clusters} and \ref{tab:n5253clusters} 
present our $EW(H\alpha)$ estimates for the cluster candidates in 
NGC~3077 and NGC~5253, along with $H\alpha$-derived age estimates.
The age estimates employ the $EW(H\alpha)$-age relations provided 
by the Starburst99 models.  As we did for the photometric ages, we 
present the plausible range of ages in addition to the best-fit age, 
by accounting for the photometric errors in the $H\alpha$ images.
In Figure~\ref{fig:ewha_ages}, we show the correlation between the 
photometric ages, and the ages derived from $EW(H\alpha)$.  The 
two age estimates are generally consistent with each other, when the 
plausible age ranges are taken into account.  

In a few cases, we found that the photometric age estimate was much 
older than the $EW(H\alpha)$ age.  We believe the most likely 
explanation for these discrepant ages is an underestimate of the 
extinction toward these clusters.  Because of the shape of the 
Starburst99 evolutionary track, it is entirely possible for a young, 
heavily-extincted cluster to be mistaken for a cluster on the red-
supergiant loop with low reddening.  However, the large $H\alpha$ 
flux associated with the young clusters reveals their true age.  
We manually apply a supplemental extinction to the photometry of 11 
clusters in NGC~3077, and 3 clusters in NGC~5253, to bring their 
intrinsic colors into agreement with their large $H\alpha$ flux (see
Tables~\ref{tab:n3077clusters} and \ref{tab:n5253clusters}).

Once we have estimated cluster ages using both broad-band photometry 
and $EW(H\alpha)$, we manually inspect the pair of plausible age 
ranges for each cluster to derive an overall best age and mass 
estimate for each cluster.  In most cases, the best photometric and 
$EW(H\alpha)$ ages are in agreement.  When they are not, we adopt 
either the best-fit photometric or $EW(H\alpha)$ age, whichever one 
falls within the plausible age range of its complement.  If neither 
best-fit age is acceptable, we choose a characteristic age from the 
intersection of the two age ranges.  When the overall best age has 
been selected, we recompute a mass for the new best age.  These 
overall best age and mass estimates are listed in 
Tables~\ref{tab:n3077clusters} and \ref{tab:n5253clusters}, for 
NGC~3077 and NGC~5253, respectively.

\subsection{Photometry of Non-Clusters}\label{sec:nonphot}

The majority of the detected objects in the broad-band images could
not be classified as clusters; we refer to these objects collectively 
as ``non-clusters'' despite the fact that the nature of many of these 
objects is indeterminate.  The extinction-corrected photometry of 243 
non-cluster objects in NGC~3077 is presented in 
Table~\ref{tab:n3077.extra}.  The photometry of 391 non-cluster 
objects in NGC~5253 is presented in Table~\ref{tab:n5253.extra}.  

33 of the non-cluster objects in NGC~3077 were rejected on the 
basis of their profile shape; the radial profiles were either 
indistinguishable from a stellar profile, or were sufficiently 
disturbed or otherwise indeterminate that no classification could be 
attempted.  Similarly, 106 of the non-cluster objects in NGC~5253 
were rejected on the basis of profile shape.  The remaining 
non-cluster objects, the majority of the total detection count in 
each galaxy, were not shape-classified because they were detected in 
only one of the three broad-band images.

In Figure~\ref{fig:partial}, we show two-color diagrams for objects 
which were undetected in either F300W or F814W (but {\it were} 
detected in the other two images).  These objects have one measured 
color and one color limit.  The color limits are determined by 
assigning the 50\% completeness magnitude to the non-detected band.  
Objects which are 1.8~mag brighter than the faint limit in F547M are 
highlighted.  These objects tend to have extremely blue 
$(m_{F547M}-m_{F814W})$ colors or extremely red 
$(m_{F300W}-m_{F547M})$ colors.  It is possible that the 
F300W-dropouts are heavily-extincted objects, but they do have rather 
blue ($m_{F547M} - m_{F814W}$) colors, so it is difficult to place 
them along the Starburst99 model, even when any arbitrary extinction 
value is addopted.

\section{Discussion}\label{sec:discuss}

\subsection{Cluster Formation Histories}\label{cfh}

The spatial distributions of our cluster-age estimates for NGC~3077 
and NGC~5253 are shown in Figures~\ref{fig:n3077.agemap} and 
\ref{fig:n5253.agemap}, respectively.  Each cluster position in the 
F547M image is labeled with its best age estimate, in Myr.  

\begin{figure}[h]
\plotone{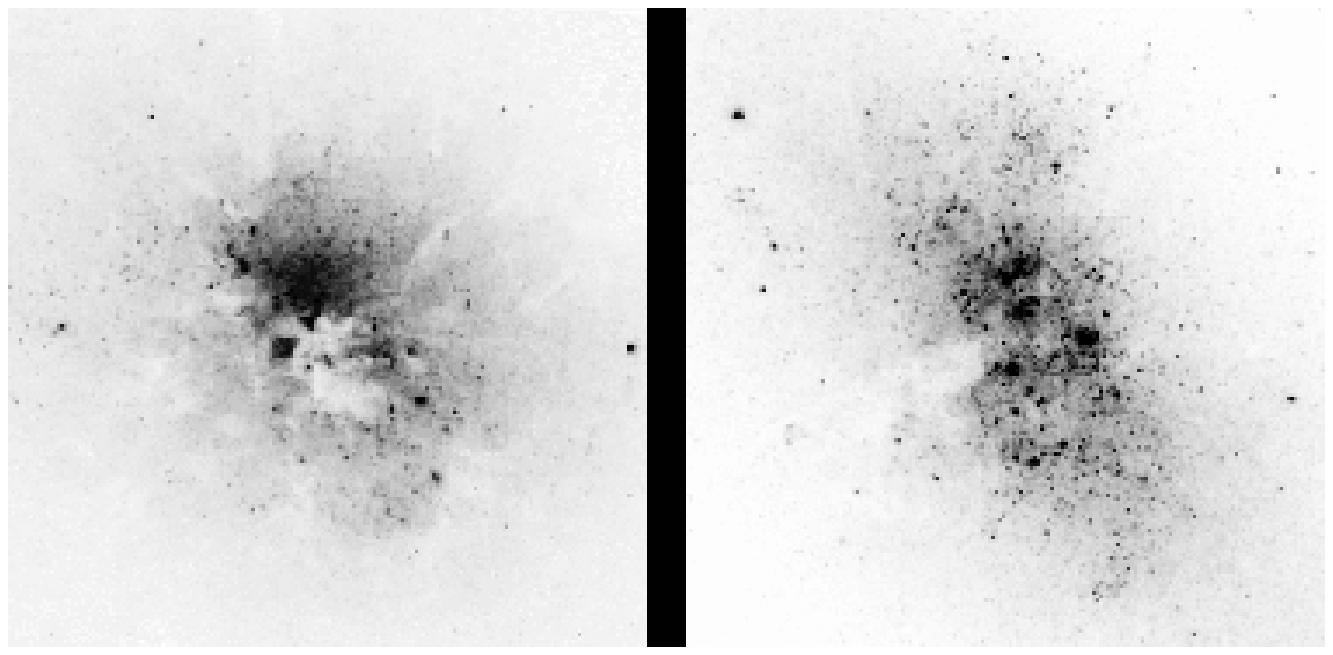}
\caption{Composite color images of NGC~3077 (left) and NGC~5253 
(right).  The RGB images were made by assigning the F300W image to 
the blue pixel values, the F547M image to the green pixel values, 
and the F814W images to the red pixel values.  In each image, North 
is at the top, and East is to the left. \label{fig:color} }
\end{figure}

\begin{figure}[h]
\plotone{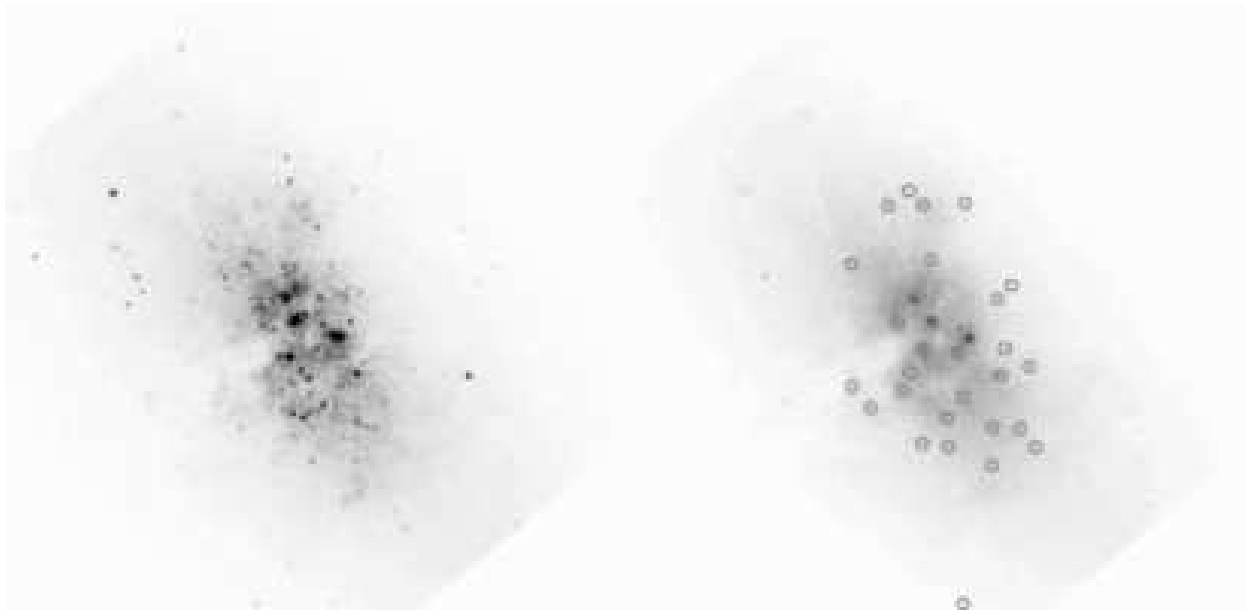}
\caption{F547M images of the central portion of NGC~5253.  The
original image is shown at left, and a sample from our series of 
neighbor-subtracted images is shown at right.  The unsubtracted 
objects are circled; all other objects have been subtracted with 
a best-fit profile model. \label{fig:subimage} }
\end{figure}

\begin{figure}[h]
\plotone{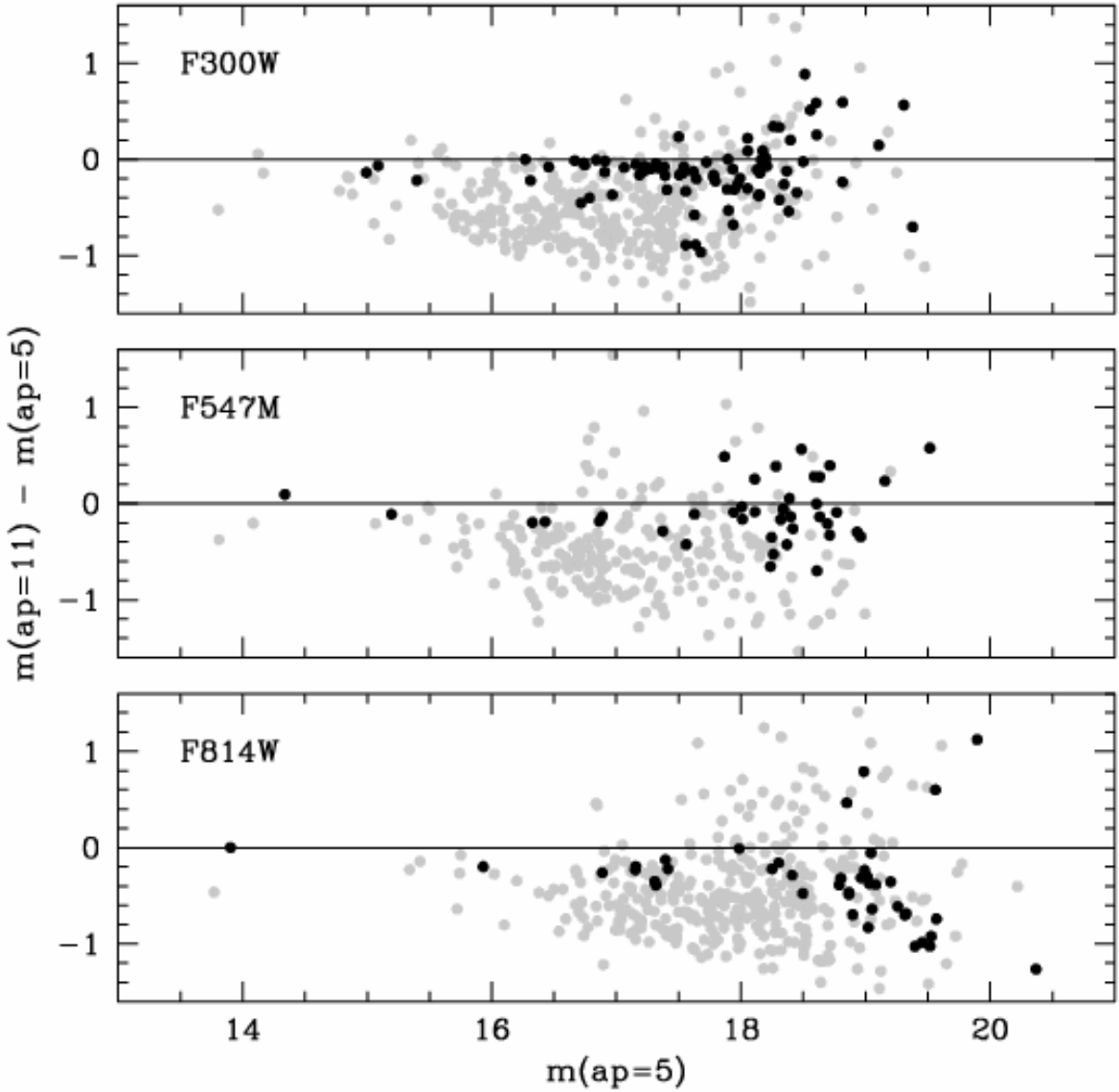}
\caption{Aperture corrections from 5 pixels to 11 pixels for the 
three broad-band filter images of NGC~3077. Each point represents 
a detected object.  Since the central region suffers from crowding 
and an elevated, variable background level, we highlight objects 
outside the central region (heavy black points).\label{fig:apcorr} }
\end{figure}

\begin{figure}[h]
\plotone{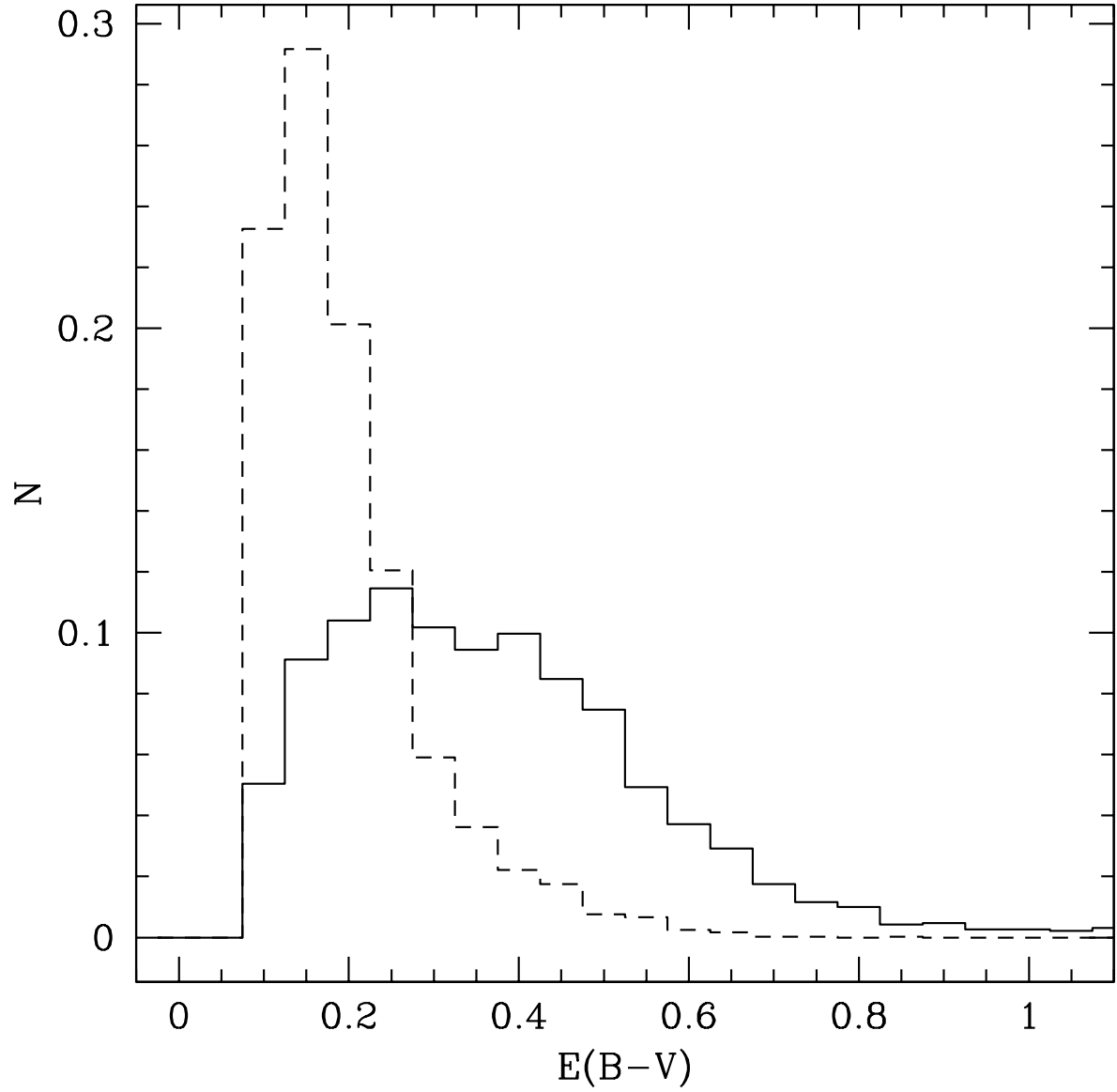}
\caption{The distribution of $E(B-V)$ reddening values in NGC~3077 
(solid histogram) and NGC~5253 (dashed histogram).  The reddening 
values are derived from a map of the $H\alpha/H\beta$ ratio in each 
galaxy. \label{fig:ebvhist} }
\end{figure}

The starburst region in NGC~3077 is dominated by a heavily-extincted 
central dust cloud (see Figure~\ref{fig:color}).  There is a large 
group of clusters younger than 10~Myr in a diffuse fan-shaped clump 
to the north of the central dust cloud.  Many of these clusters 
appear to be nearly coeval, with ages between 5 and 7~Myr.  The most 
massive of these is strikingly apparent as a blue knot to the 
northeast of the dusty core in Figure~\ref{fig:color} (object~\#12 in 
Table~\ref{tab:n3077clusters}).
 
The youngest cluster in our NGC~3077 sample (aged 1~Myr) lies inside 
the central dust cloud (object~\#43 in
Table~\ref{tab:n3077clusters}); there are also two objects nearby 
that are so heavily extincted that they were undetected in the 
F300W image; these objects are likely also very young clusters that 
have yet to emerge from the surrounding dust cloud.  

There are several clusters in NGC~3077 aged between 10 and 20~Myr, 
and these are scattered throughout the central starburst region.  The 
visually brightest cluster in our NGC~3077 sample falls in this age 
range (object~\#3 in Table~\ref{tab:n3077clusters}, aged 13~Myr), and 
is located adjacent to the central dust cloud.  This cluster's heavy 
extinction ($E(B-V)=0.7$~mag) and close proximity to the central dust 
cloud suggest that its age may be overestimated.  Indeed, if we 
adopted an even larger extinction, it would be possible to place the 
cluster on the youngest portion of the Starburst99 model track in 
Figure~\ref{fig:2cd}.  However, the cluster has a very low 
$EW(H\alpha)$ (see Table~\ref{tab:n3077clusters}), making it more 
likely that its present age estimate is the correct one.  NGC~3077 
contains several clusters older than 20~Myr, up to 300~Myr old.  Most 
of these are located to the west and southwest of the central dust 
cloud, and are relatively far from the galaxy's center.

In NGC~5253, we see a tight group of very young (age $<5$~Myr) 
clusters near the center of the galaxy, and there are similarly young 
clusters scattered throughout the central starburst region, including 
the "super star cluster candidate" of \cite{cal97b}; object \#1 in 
Table~\ref{tab:n5253clusters}.

In contrast to NGC~3077, the clusters with ages between 5 and 10~Myr 
in NGC~5253 are scattered throughout the starburst region, with no 
obvious concentrations.  Looking at slightly older objects, there are 
two very bright clusters or groups of clusters in NGC~5253 that have 
ages around 10~Myr.

The cluster populations in NGC~3077 and NGC~5253 suggest that star 
formation has proceeded in discrete, highly-correlated regions of 
activity producing tight groups of clusters which dissipate their 
parent molecular cloud on a timescale of a few~Myr.  However, there 
appears to be a more diffuse component as well, since some of the 
very young clusters are isolated.  The clusters themselves are 
probably not bound, and will likely dissipate on a short timescale 
(see Section~\ref{sec:dissolve}).

In Figure~\ref{fig:mass-age}, we plot our best estimate of each 
cluster's age against its best mass estimate.  The curves in 
Figure~\ref{fig:mass-age} represent the mass corresponding to the 
90\% completeness limit as a function of cluster age.  To construct 
this curve, we first determine the 90\% completeness magnitude, 
$m_{90}$, in each band using artificial object tests.  We then scale 
the mass of each model cluster along the Starburst99 track until its
magnitude is $m_{90}$.  The artificial objects are generated using 
the best-fit cluster profiles which we used to subtract objects for 
our series of neighbor-cleaned images (see 
Section~\ref{sec:analysis}).  Artificial clusters are distributed 
randomly in each image, but only within a central subregion where 
most of the real clusters are found.  The completeness curves indicate 
that evolutionary fading biases our sample against old, low-mass 
clusters: roughly half of the young (age $<20$~Myr) clusters would be 
unobservable if they were 100-200~Myr old.  When the possiblility of 
mass-loss by evaporation is considered, the bias with age becomes 
even more significant.  Figure~\ref{fig:mass-age} also indicates that 
while NGC~3077 contains many more clusters than NGC~5253, the 
distribution of cluster masses appears to be similar in both 
galaxies.

We plot the distribution of cluster ages in Figure~\ref{fig:agehist}.
The shaded histograms are the ages of clusters brighter than 
M$_V=-9$~mag.  Both galaxies show a strong peak of clusters with ages 
between 1 and 20~Myr, but the peak in NGC~3077 is much sharper.  It 
has around 20 clusters aged 5--7~Myr, while NGC~5253 has only a few 
clusters in this age range.  NGC~3077 has a number of clusters older 
than 20~Myr, but there are no clusters in this age range in NGC~5253.

\subsection{The Interstellar Medium}\label{ism} 

\begin{figure}[h]
\plotone{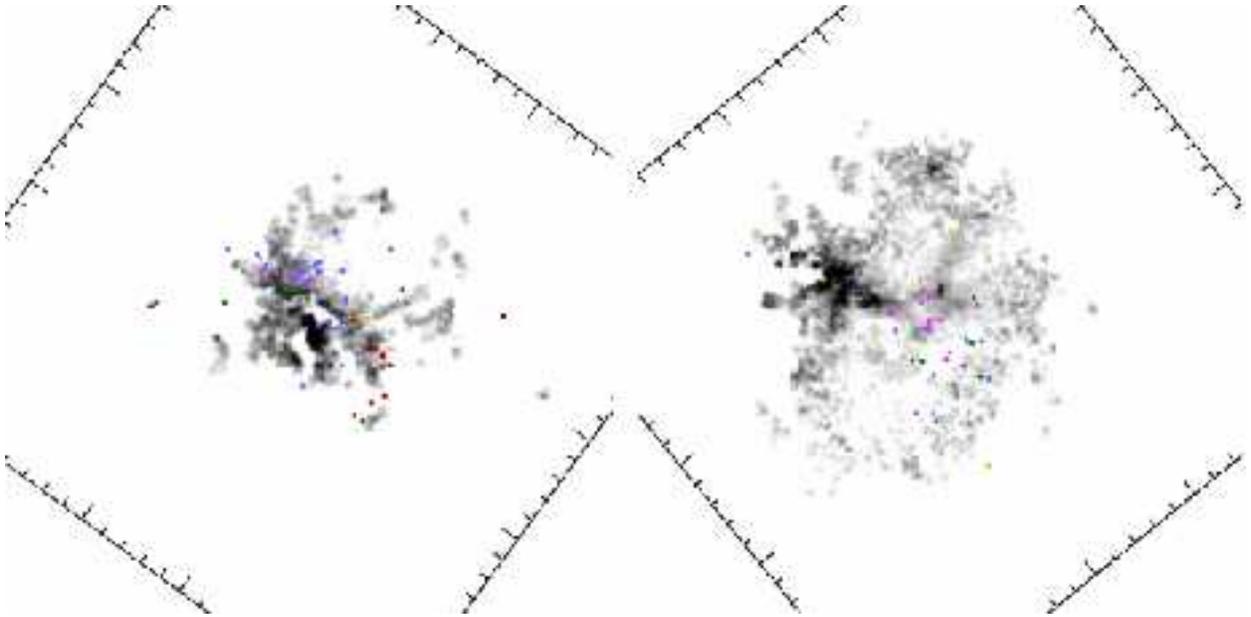}
\caption{Maps of the $H\alpha/H\beta$-derived E(B-V) values in 
NGC~3077 (left) and NGC~5253 (right).  North is at the top, East is 
at the left.  White pixels have $E(B-V)=0.0$~mag; black pixels have 
$E(B-V)=1.0$~mag. The positions of our confirmed cluster candidates 
are indicated with points.  In the electronic edition of the paper, 
the cluster points are color-coded according to the age estimate of 
the cluster, as in Figure~\ref{fig:n3077.clusterid}. 
\label{fig:ebvmap} }
\end{figure}

\begin{figure}[h]
\plotone{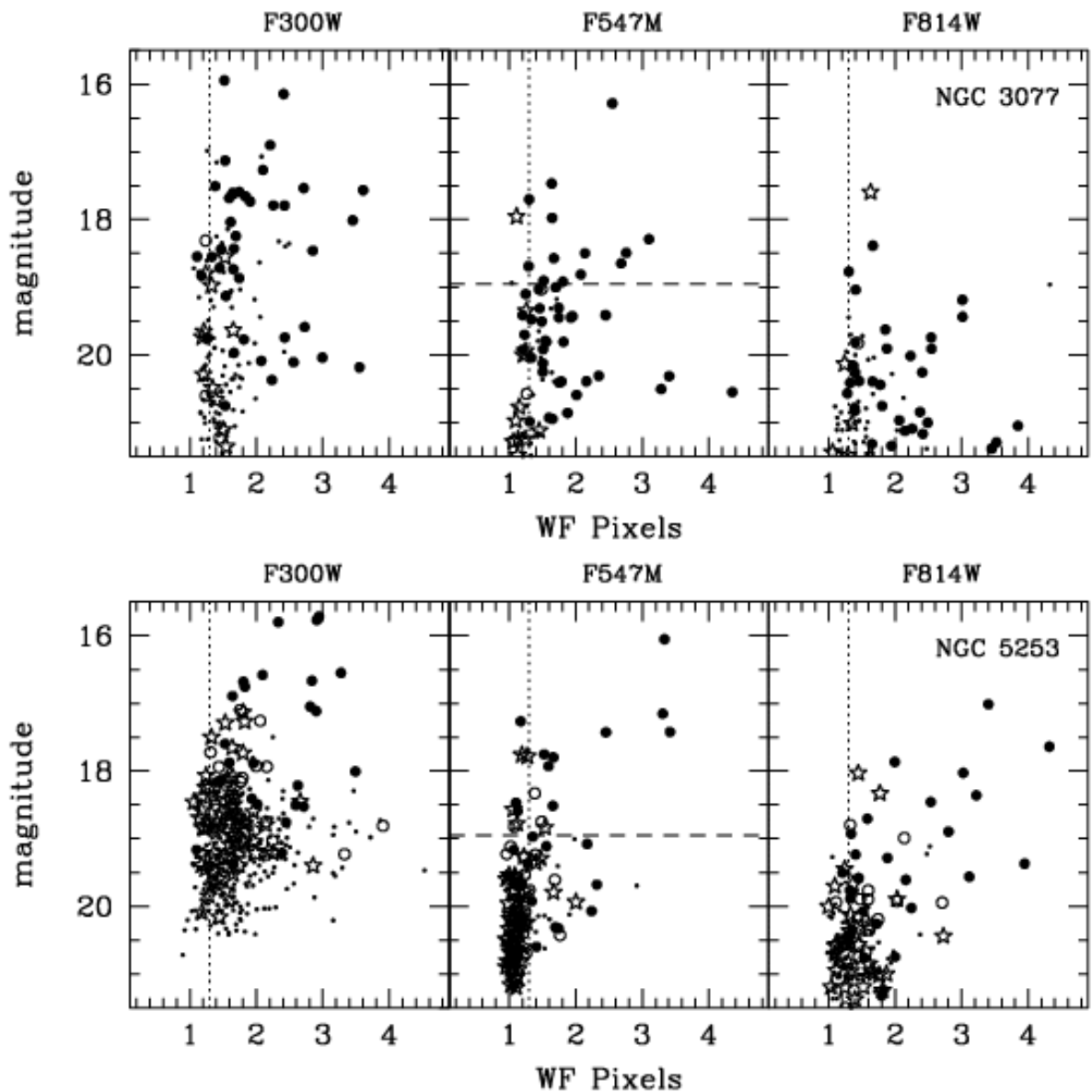}
\caption{The FWHM measurements, in WF pixels, for all detected 
objects, as a function of magnitude for the F300W (left column), 
F547M (center column), and F814W (right column) images.  The top 
row shows objects in NGC~3077, while the bottom row shows 
NGC~5253's objects.  In each panel, objects which were determined 
to be clusters are shown as solid points, objects which were 
determined to be stars are shown as open stars, and ambiguous 
objects are shown as open circles.  Objects which were detected in 
only one filter are shown as small points.  The instrumental PSF size 
is shown as a vertical dotted line, and $M_V=-9.0$~mag is indicated 
with a dashed line in the center panels. \label{fig:fwhm} }
\end{figure}

\begin{figure}[h]
\plotone{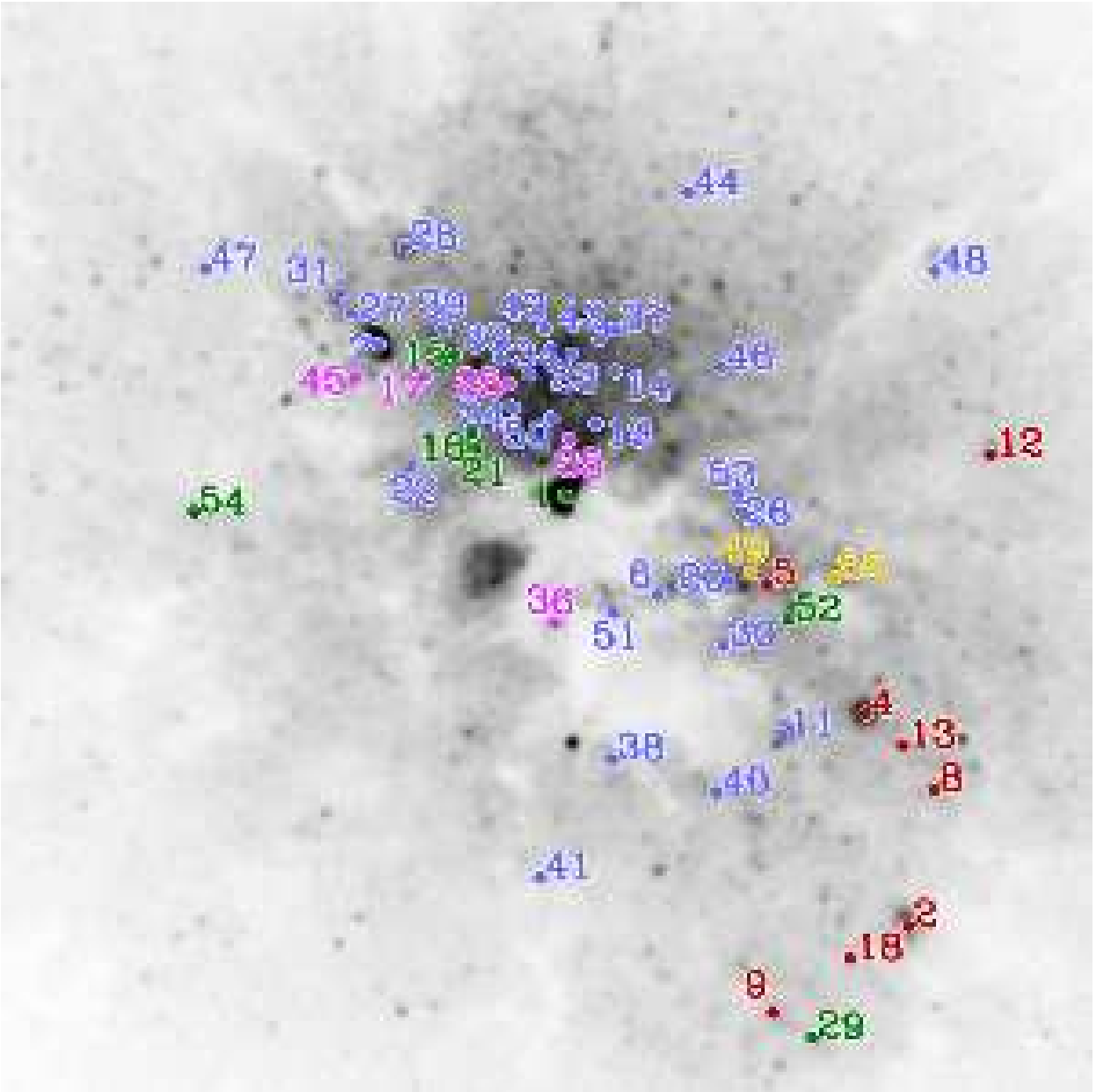}
\caption{The positions of our confirmed cluster candidates in 
NGC~3077, superimposed on the F547M image (oriented such that North 
is at the top, East is to the left). Each cluster's position is 
labeled with its ID number from Table~\ref{tab:n3077clusters}.  The 
ID numbers are color-coded according to the age estimate of the 
cluster: 1--5~Myr (violet), 6--10~Myr (blue), 10--20~Myr (green), 
20--40~Myr (orange), 40--1000~Myr (red).
\label{fig:n3077.clusterid} }
\end{figure}

\begin{figure}[h]
\plotone{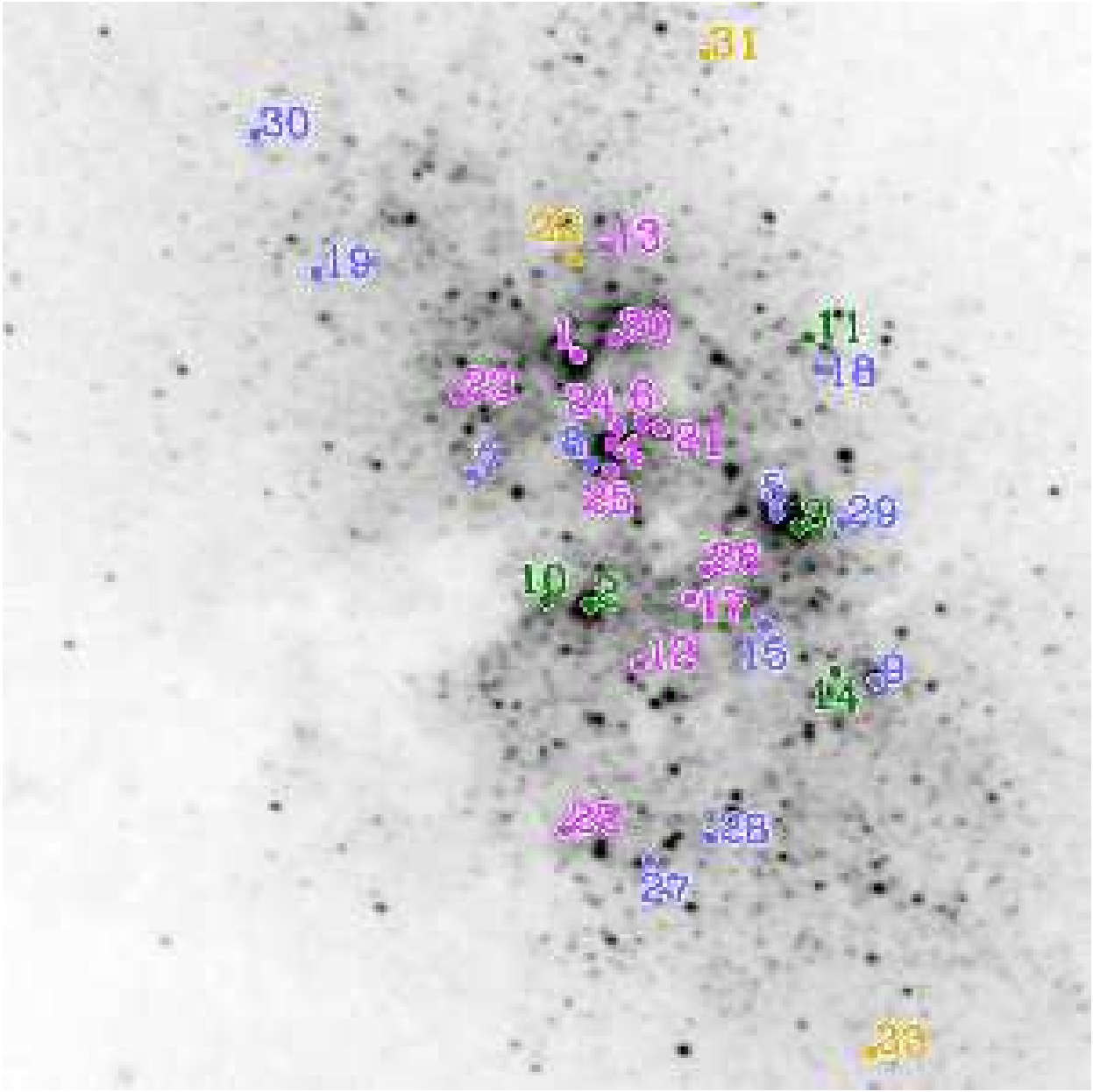}
\caption{The positions of our confirmed cluster candidates in 
NGC~5253, overplotted on the F547M image (oriented such that North 
is at the top, East is to the left).  Each cluster's position is 
labeled with its ID number from Table~\ref{tab:n5253clusters}.  The 
ID numbers are color-coded according to the age estimate of the 
cluster, as in Figure~\ref{fig:n3077.clusterid}. 
\label{fig:n5253.clusterid} }
\end{figure}

\begin{figure}[h]
\plotone{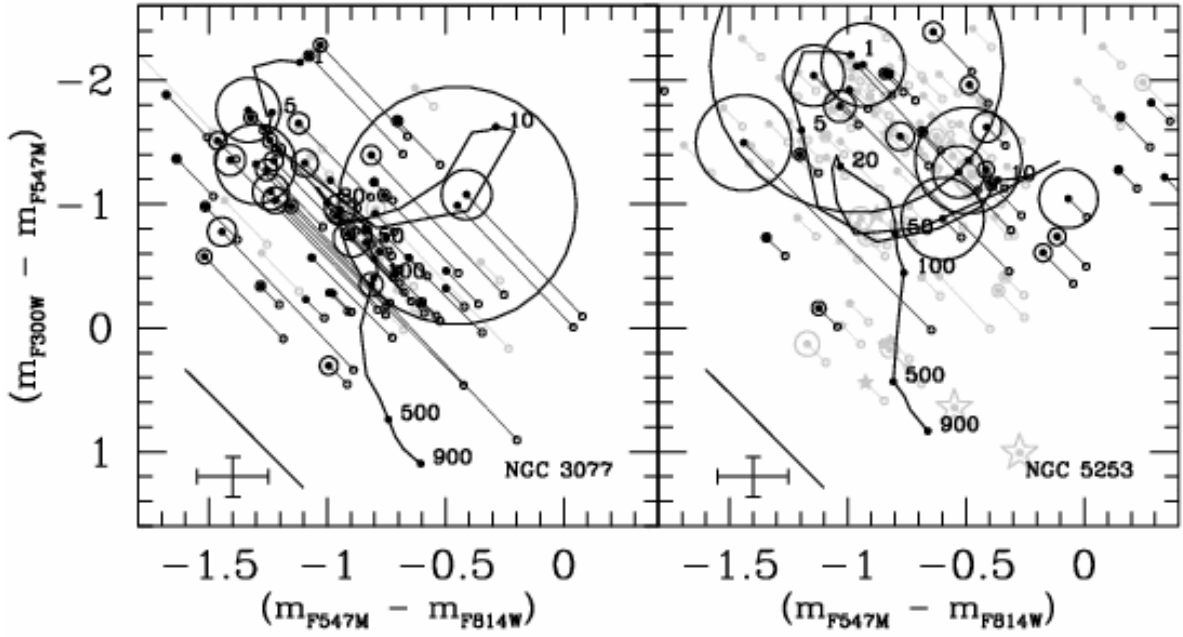}
\caption{The two-color diagram for our cluster candidates in 
NGC~3077 (left panel) and NGC~5253 (right panel).  Small open points 
represent the observed photometry.  Solid points represent the 
intrinsic photometry, after de-reddening.  Each observed/intrinsic 
pair is connected with a reddening line.  In addition, a circle whose 
size is proportional to the object's F547M flux is superimposed on 
the intrinsic photometry point.  The confirmed clusters are shown as 
black circles; confirmed stellar objects are shown as grey stars; 
ambiguous objects are shown as grey circles.  The solid black curves 
are Starburst99 model tracks with an instantaneous-burst 
star-formation rate and an appropriate metallicity for each galaxy 
(solar metallicity for NGC~3077; $\frac{1}{4}$ solar for NGC~5253). 
\label{fig:2cd} }
\end{figure}

\begin{figure}[h]
\plotone{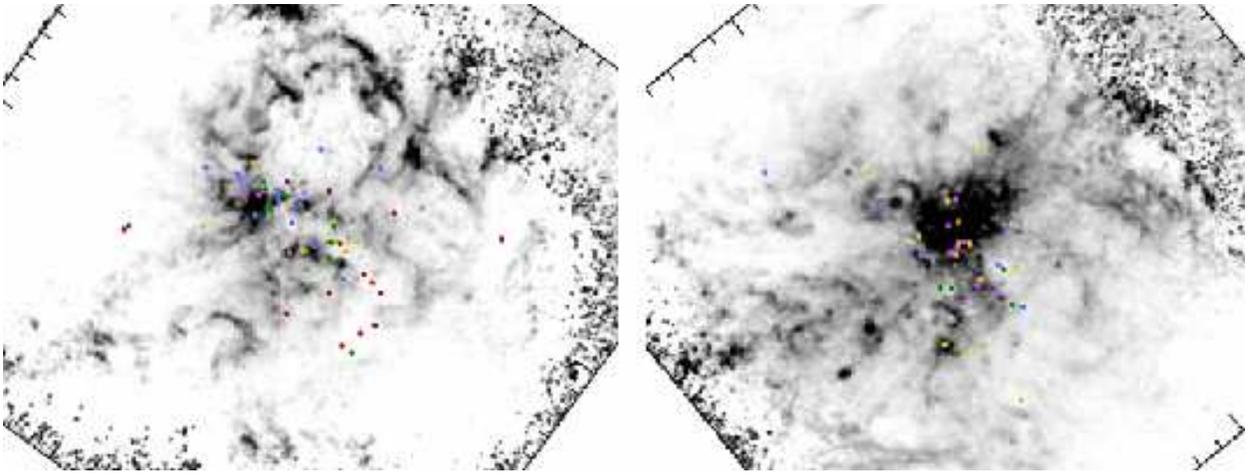}
\caption{$EW(H\alpha)$ maps for NGC~3077 (left) and NGC~5253 (right).  
White corresponds to $EW(H\alpha)=0$\AA\ in both images.  Black 
corresponds to $EW(H\alpha)=250$\AA\ in NGC~3077, and 
$EW(H\alpha)=1000$\AA\ in NGC~5253.  In each image, North is at the 
top, East is to the left.  The positions of our confirmed cluster 
candidates are indicated with points.  In the electronic edition of 
the paper, the points are color-coded according to the age 
estimate for each cluster, as in Figure~\ref{fig:n3077.clusterid}. 
The noisy fringe in the outer regions of both images is an artifact 
due to low signal-to-noise. \label{fig:halpha} }
\end{figure}

\begin{figure}[h]
\plotone{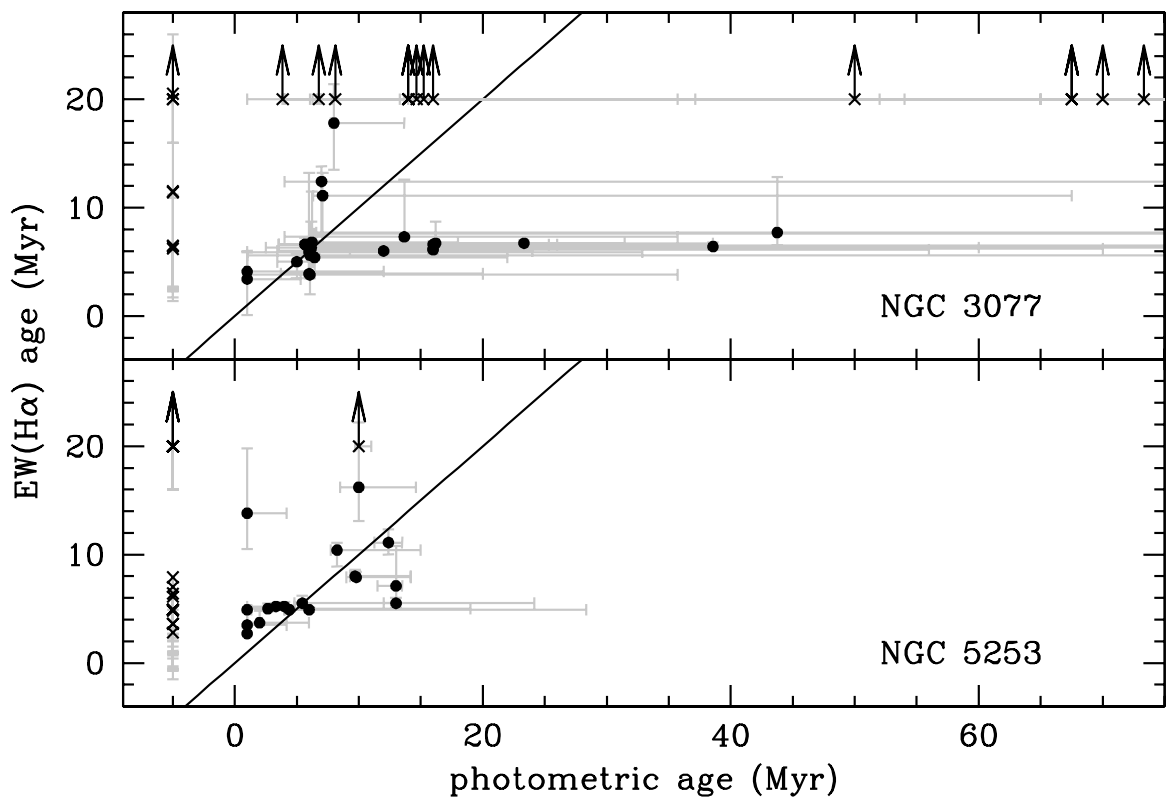}
\caption{The correlation between cluster age estimates derived from 
broad-band colors and those derived from $EW(H\alpha)$.  The range of 
plausible ages are indicated for each measurement with an error bar. 
The diagonal line indicates the locus for which the two ages are 
equal. Filled circles indicate clusters with valid age estimates from 
both the broad-band colors and $EW(H\alpha)$.  Cross symbols indicate 
clusters which have either no photometric age estimate (plotted at 
$t_{phot}=-5~Myr$) or no detectable $EW(H\alpha)$ (lower limits at 
$t_{EW(H\alpha)}=20~Myr$). 
\label{fig:ewha_ages} }
\end{figure}

\begin{figure}[h]
\plotone{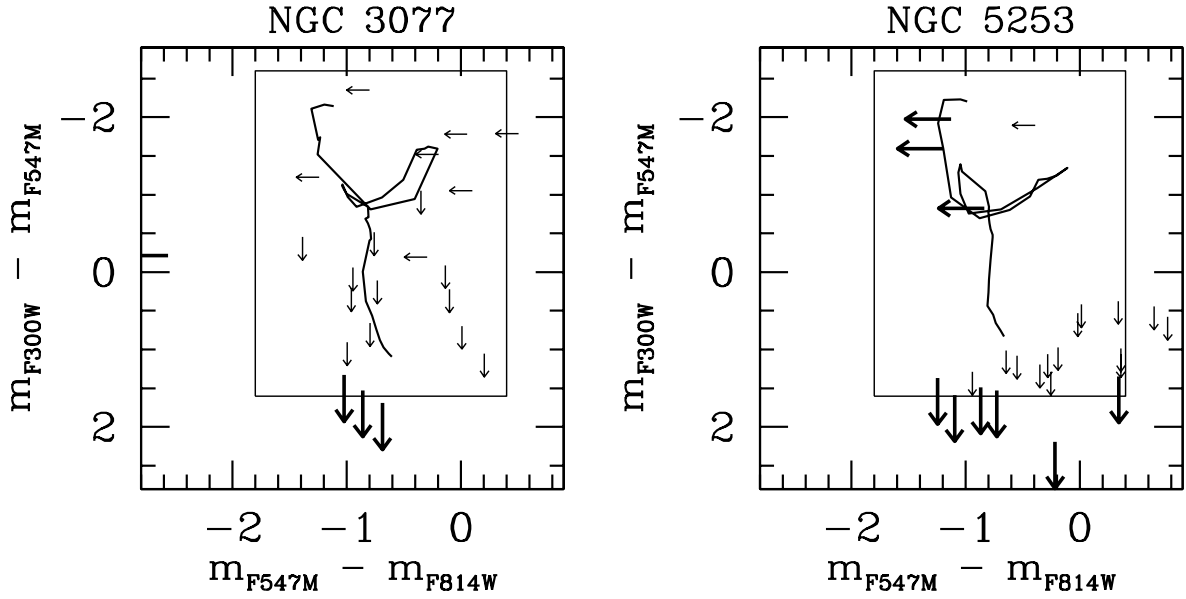}
\caption{The extinction-corrected photometry of objects which were 
detected in only two of the three broad-band filters.  For these 
objects, we have only one measured color, and a limit on the other 
color.  The objects which were undetected in F814W are shown as 
left-pointing arrows, and the objects which were undetected in F300W 
are shown as down-pointing arrows.  The inset box in each panel 
indicates the limits of the two-color diagrams plotted in 
Figure~\ref{fig:2cd}. Objects which are more than 2 magnitudes 
brighter than the 50\% completeness limit in F547M are shown with 
larger arrow symbols.  The color limits on these objects imply 
extreme colors. \label{fig:partial} }
\end{figure}

\begin{figure}[h]
\plotone{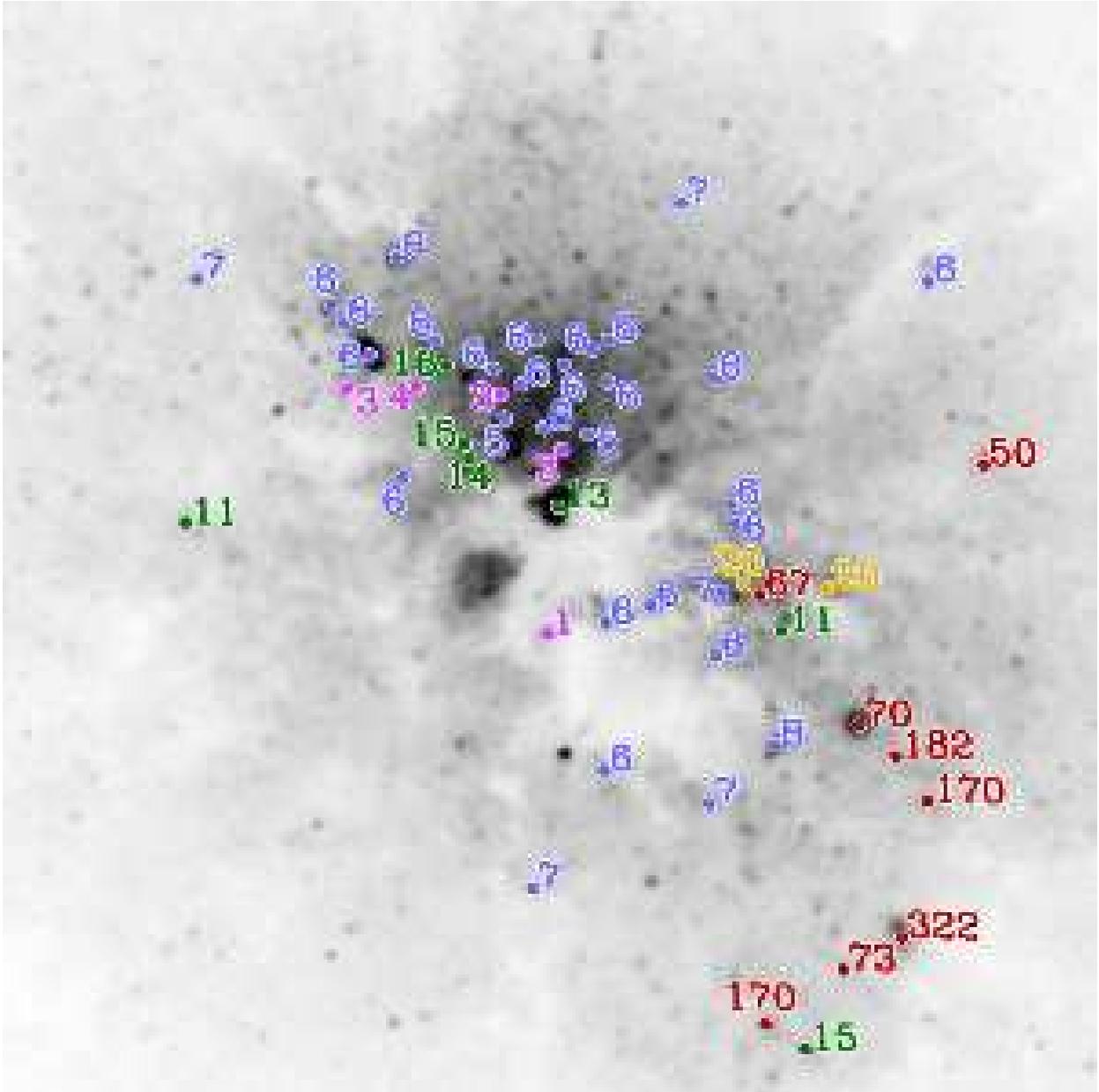}
\caption{The distribution of cluster ages in NGC~3077, superimposed 
on the F547M image (oriented such that North is at the top, East is 
to the left).  Each cluster position is labeled with a number 
indicating its age, in Myr.  In the electronic edition of the paper, 
the age labels are color-coded according to their age range, as in 
Figure~\ref{fig:n3077.clusterid}. \label{fig:n3077.agemap} }
\end{figure}

\begin{figure}[h]
\plotone{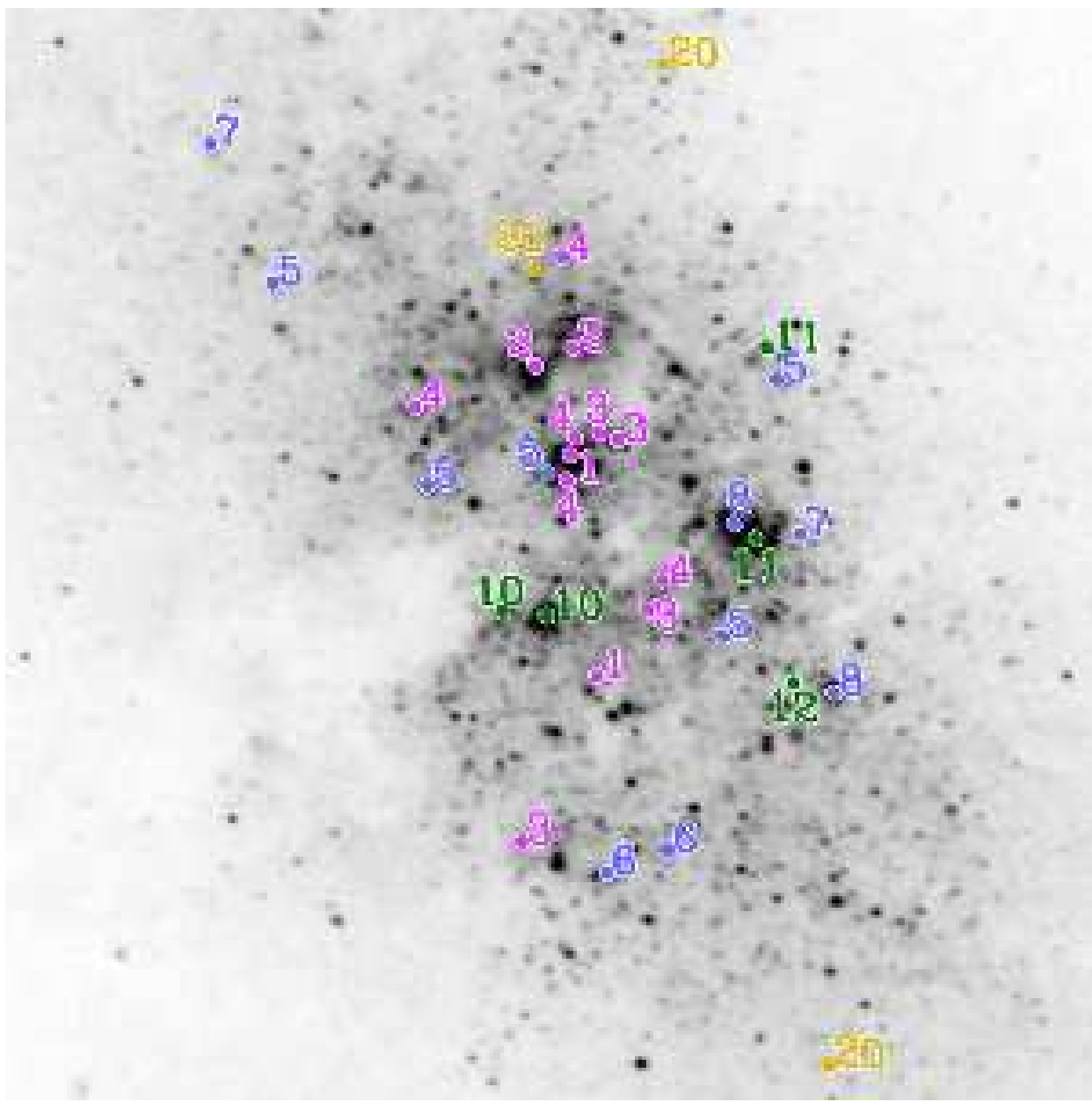}
\caption{Same as Figure~\ref{fig:n3077.agemap}, for clusters in 
NGC~5253. \label{fig:n5253.agemap} }
\end{figure}

\begin{figure}[h]
\plotone{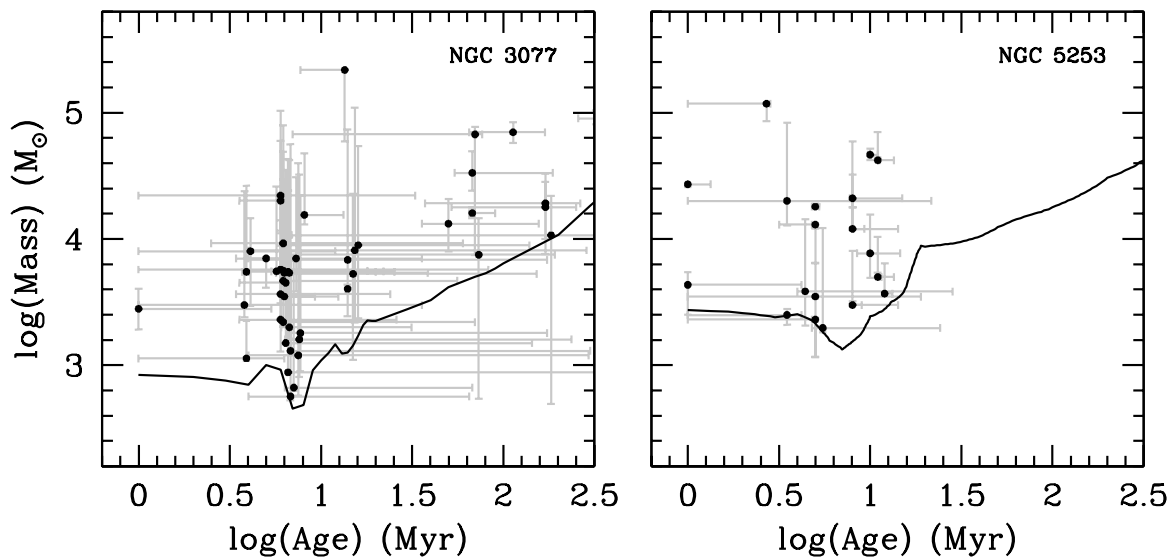}
\caption{The relationship between mass and age for the cluster
candidates in NGC~3077 (left panels) and NGC~5253 (right panels).  
The points represent our best age and mass estimates for each 
cluster, while errorbars indicate the plausible range of age and 
mass.  The curves represent the 90\% completeness limit, transformed 
to mass.  The top row plots the ages and masses on linear axes, while 
the bottom row uses logarithmic axes. \label{fig:mass-age} }
\end{figure}

\begin{figure}[h]
\plotone{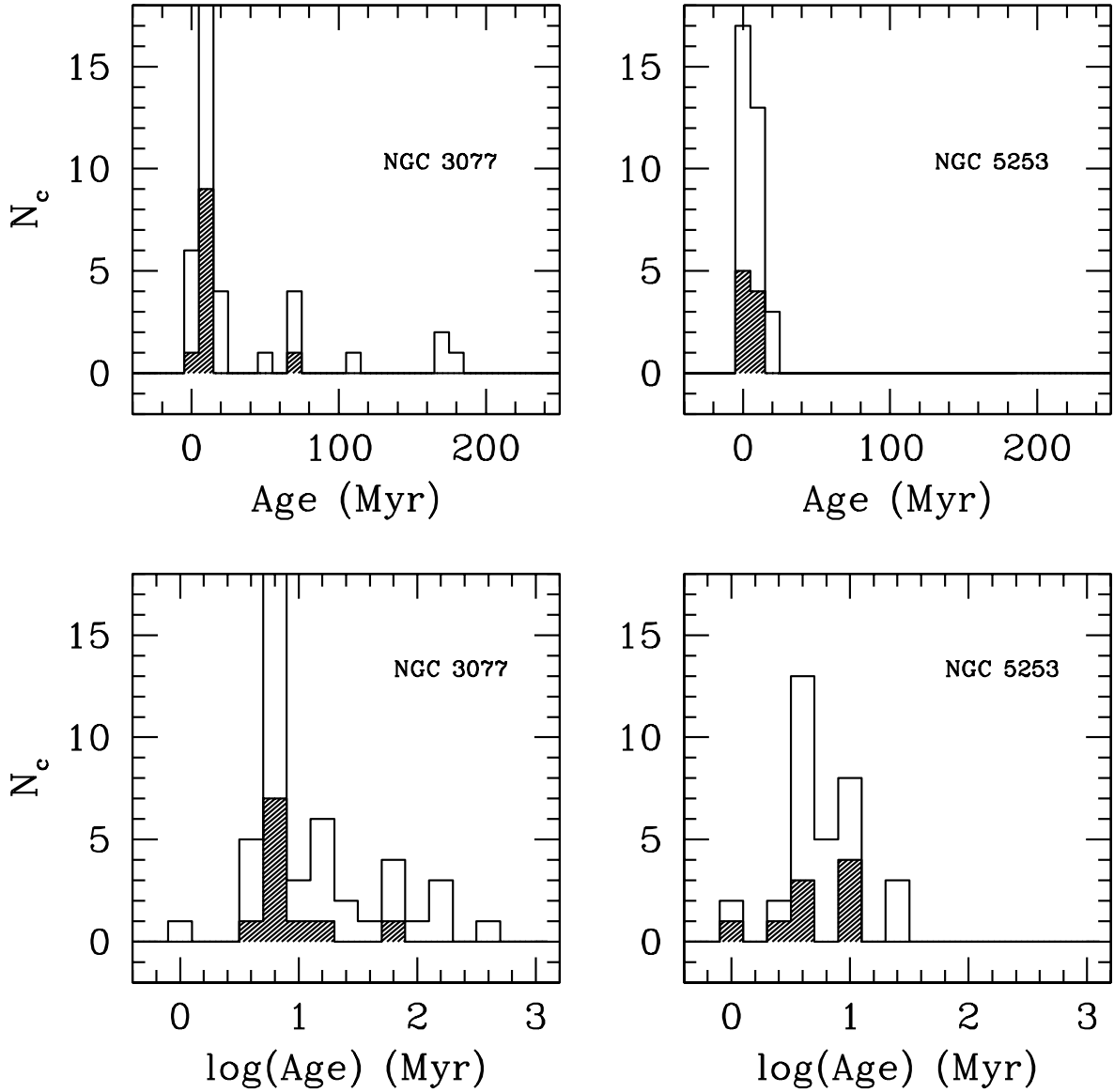}
\caption{The distribution of cluster ages for NGC~3077 (left panels) 
and NGC~5253 (right panels).  The top row plots the ages linearly; 
the bottom row plots the ages logarithmically. Shaded histograms show 
the distribution of ages for the subset of clusters with 
$M_V < -9$~mag. \label{fig:agehist} }
\end{figure}

NGC~3077 and NGC~5253 exhibit some interesting contrasts in the 
relationship between their cluster populations and their interstellar
gas and dust content.  NGC~3077 appears to be a dustier galaxy 
overall, as can be seen in Figures~\ref{fig:color}, \ref{fig:ebvhist} 
and \ref{fig:ebvmap} (the two regions of apparently-zero extinction 
at the center of NGC~3077 in Figure~\ref{fig:ebvmap} are actually 
extremely extincted; the $H\beta$ flux is too low in these regions to 
obtain an extinction estimate).  Furthermore, Figure~\ref{fig:ebvmap} 
suggests that the clusters in NGC~3077 are more likely to be found in 
substantially extincted regions, compared to the clusters in 
NGC~5253.  

The distribution of $H\alpha$-emitting gas in these galaxies 
(Figure~\ref{fig:halpha}) reveals more contrasts between these 
galaxies.  The $H\alpha$ gas in NGC~3077 exhibits a shell/bubble 
morphology \citep{mar98}, and the largest peak in its $H\alpha$ is 
adjacent to (but not coincident with) the location of a grouping of 
several young clusters.  NGC~5253 has much more $H\alpha$ emission 
overall (note the different greyscales used in the panels of 
Figure~\ref{fig:halpha}), and it is strongly concentrated into a 
single, dominant clump of $H\alpha$ gas.  The super star cluster 
candidate of \cite{cal97b} lies at the very center of this $H\alpha$ 
concentration (object \#1 in Table~\ref{tab:n5253clusters}).

The larger $H\alpha$ flux from NGC~5253 suggests a larger population 
of very young stars.  Yet we actually observe a {\it smaller} 
population of young clusters in NGC~5253.  Since NGC~5253 contains 
many more objects which may be individual stars in a diffuse field 
component, one might expect that a substantial portion of the  
$H\alpha$ may have been ionized by this diffuse population.  However, 
\cite{tre01} concluded from long-slit UV STIS spectra that the 
intercluster field in NGC~5253 is {\it devoid} of the massive O stars 
that are largely responible for photoionization.  NGC~5253 contains 
several $H\alpha$ ``hotspots'', peaks that have no optical 
counterpart whatsoever.  These hotspots are likely extremely young 
clusters that have yet to emerge from the molecular cloud complexes 
from which they formed.  These embedded young clusters may explain 
the apparent overabundance of $H\alpha$ in NGC~5253.

\subsection{Cluster Dissolution Timescales and the Diffuse Field Population}\label{sec:dissolve}

Determining cluster dissolution timescales and studying the field 
stellar populations of starburst galaxies can provide constraints on 
important questions regarding the origin of field populations.  Are 
the field populations of galaxies composed of dissolved clusters, or 
did they form {\it in situ}, from a diffuse field mode of star 
formation?  Is the origin of the field population different for 
different galaxies?

In NGC~5253, we observe a complete lack of clusters older than 
$\sim20$~Myr, while NGC~3077 contains many such clusters, and 
several which are older than 100~Myr.  This contrast cannot be 
entirely understood as a difference in the completeness rates in the
two galaxies.  Applying NGC~5253's completeness curve to NGC~3077's
cluster population in Figure~\ref{fig:mass-age} results in seven 
clusters older than 10~Myr becoming unobservable, which is less than 
half the total number of its clusters in this age range.  Does the 
lack of older clusters in NGC~5253 indicate that cluster dissolution 
is more efficent in NGC~5253, or simply that the star-formation 
episode in NGC~5253 began only $\sim20$~Myr ago?  We examine the 
dynamic environments of the two galaxies to address this question.

As \cite{tre01} already determined, evaporation timescales for 
clusters in the center of NGC~5253 are around 20~Myr, using the
figures of \cite{kml99} for the Milky Way rescaled to the case of 
NGC~5253. Although NGC~5253 is about two orders of magnitudes less 
massive than the Milky Way, the evaporation timescales for stellar 
clusters within $\sim$100-150~pc from its center are only 
$\sim$4~times longer than for our Galaxy under the same conditions. 
The main reason resides in the large velocity dispersion measured by 
\cite{cp89}, $\sigma\sim$50~km~s$^{-1}$, in the center of the dwarf 
galaxy, much larger than for instance, its central rotation velocity 
(a few~km~s$^{-1}$).  By applying the virial theorem to this central 
dispersion velocity, we obtain an estimate for the mass contained 
within the central 100~pc of NGC~5253, which is only $\sim$15~times 
smaller than the mass contained within the same radius in the Milky 
Way \citep{kml99}.  This suprisingly large central mass concentration 
leads to small predicted evaporation timescales for the star clusters 
in the center of NGC~5253. For the range of cluster 
masses we detect, the evaporation timescales are in the range 
16--50~Myr. These values may be uncertain by as much as a factor of 
two, because of the uncertainty in the techniques used to model the 
evolution of clusters \citep[\eg\ ][]{por02}. In addition, mass 
segregation can contribute to accelerate tidal evaporation of 
clusters \citep{kml99}.  Combining the cluster dissolution 
information with the lack of massive stellar populations in the 
intercluster field of NGC~5253 \citep{tre01} suggests that most stars 
in NGC~5253 form in star clusters, which then dissolve into the 
field population on a timescale of $\sim20$~Myr.  The two large 
clusters in NGC~5253 which are $\sim10$~Myr old are interesting in 
this context; perhaps these objects are massive enough to survive the 
harsh dynamical environment of NGC~5253's core.

NGC~3077 appears to have a comparable rotation velocity to that of 
NGC~5253, but a much smaller central dispersion velocity; direct 
measurements from broadening of stellar features are not available 
for this galaxy; the small dispersion is inferred from molecular 
clouds velocities \citep{wal02}. If these velocities are 
representative, the central dispersion in this galaxy is 
$<$20~km~s$^{-1}$, or about a factor 2.5 smaller than in NGC~5253. 
Because of the strong tidal fields in the environments of the
cluster populations, we associate the evaporation timescale of star 
clusters with the dynamical time of the star cluster's orbit
in the host galaxy \citep[see Eq. 2 in\ ][]{kml99}.  Under this 
assumption, the difference in central velocity dispersions 
implies that cluster evaporation timescales are 2.5 times longer
in NGC~3077, compared to NGC~5253, which can perhaps 
account for the older clusters present in NGC~3077.

\subsection{Comparing to the Starburst in M~83}\label{sec:m83}

It is instructive to compare the nature of the starbursts in the 
dwarf galaxies NGC~3077 and NGC~5253 to the nuclear starburst of 
the grand-design spiral galaxy, M~83.  While the three starburst 
regions contain similar total numbers of clusters, the M~83 starburst 
has many more bright clusters (33 clusters with $M_V < -9$~mag); than 
NGC~3077 or NGC~5253, which have 12 and 11, respectively.  Since M~83 
is 5--10 times more massive than these dwarf galaxies, it is perhaps 
not suprising that its starburst contains a proportionately larger 
number of massive star clusters.  However, while the mass 
distribution is more top-heavy in M~83, we note that the mass of the 
largest star clusters in all three galaxies is about the same: 
$\sim10^5 M_\odot$ (see Figure~\ref{fig:mass-age} and Figure~11 in 
\citet{har01}).

Furthermore, we find no correlation between the physical size of the 
starburst region and the size of the galaxy.  In all three galaxies, 
the starburst regions (as defined by the presence of young clusters) 
span $\sim300$~pc.  In M~83, this amounts to only the innermost few 
percent of the galaxy's total radial extent; while in NGC~3077 and 
NGC~5253, the starburst regions encompass a much larger fraction of 
each galaxy's optical extent.  This contrast in the starburst's 
significance to the whole galaxy may point to a fundamental 
difference in the nature of the starbursts.  In M~83 it is likely 
that gas and dust are funneled into the nuclear region by the 
galaxy's bar instability, creating a localized region overdense 
with prestellar material, and ripe for intense star formation.  
However, these dynamical conditions do not exist in the dwarf 
galaxies NGC~3077 and NGC~5253, so the conditions leading them to 
starburst may be fundamentally different.  

Another interesting contrast between M~83 and these dwarf galaxies is 
in the spatial distribution of their youngest clusters.  In 
\citet{har01}, we noted the presence of an ``active ringlet''; a 
semicircular annulus centered on the optical nucleus of the galaxy.  
This ringlet contains the bulk of the young clusters in M~83's 
starburst, and also appears to be a hot cavity of ionized 
interstellar material, surrounded by denser gas.  The young clusters 
span a very narrow range in age (5--10~Myr), and appear to be 
more-or-less uniformly distributed throughout the ringlet.  There is 
some evidence that a handful of even younger clusters are distributed 
along the rim of the cavity, perhaps partially embedded in the 
surrounding denser gas.  The history implied by this evidence is of a 
highly-correlated star formation event that formed most of the 
clusters in the ringlet at about the same time.  In NGC~3077 and 
NGC~5253, on the other hand, we see starburst regions that are not so 
well-defined, and cluster populations spanning a slightly wider range 
of ages (1--20~Myr).  In addition, we see discrete groups of 
coeval star clusters, distinctly younger than the ages of the 
more diffusely distributed clusters.  The history implied is one 
where cluster formation is more stochastic in these dwarf galaxies; 
small groups of clusters form together in the densest regions of the 
interstellar material, which is then dispersed by the energy output 
of the clusters after only a few~Myr.

\section{Summary}\label{sec:summary}

We present {\it HST} photometry of stars and star clusters in the 
nearby dwarf starburst galaxies NGC~3077 and NGC~5253.  We use 
standard aperture photometry in our analysis, but the photometry 
is performed on images in which nearby neighbors have been subtracted
using a best-fit profile model.  We separate star clusters from 
individual stars using the profile-shape information provided by 
the WFPC2 images.  We have measured photometry in the F300W, F547M, 
nd F814W bands for 55 star clusters in NGC~3077 and 33 star clusters
in NGC~5253.  The photometry is extinction-corrected using the ratio
of $H\alpha$ to $H\beta$.  

By comparing the extinction-corrected cluster photometry to 
Starburst99 population synthesis models, we derive age and mass 
estimates for each cluster.  We also measure $EW(H\alpha)$ toward
each cluster, which provides an additional age constraint for 
young clusters.  Both galaxies have abundant populations of star 
clusters with ages less than 20~Myr, and masses between a few 
thousand and $10^5 M_\odot$.  

We discuss some interesting contrasts between these two cluster 
populations.  NGC~3077 has more clusters overall, and about half of
them lie in the very narrow age range between 5 and 7~Myr.  These 
coeval clusters are distributed in a loose, fan-shaped region to the 
northeast of the central dust cloud in NGC~3077.  NGC~5253 has a 
large number of very young (age $<5$~Myr) star clusters, and about 
half of these are in one tight clump near the center of the galaxy.
NGC~3077 has only five clusters younger than 5~Myr, and they don't 
appear to be spatially correlated.  However, the youngest cluster in 
NGC~3077 lies within the central dust cloud, and there is some 
evidence for other young objects nearby that have yet to fully emerge 
from the dusty core.  NGC~3077 contains several intermediate-age 
clusters, up to several hundred~Myr old (although all clusters older 
than $\sim20$~Myr are relatively far from the galaxy's center).  In 
contrast, NGC~5253 contains no clusters older than $\sim20$~Myr.  
This difference is possibly due to the different cluster dissolution 
timescales in these two galaxies.  We estimate that typical 
dissolution timescales are $\sim2.5$ times longer in NGC~3077 than in 
NGC~5253.

The star cluster populations suggest that star formation has 
proceeded in discrete, highly-correlated clumps in these galaxies, 
and that this initial structure is dissipated on a short timescale 
(of order 20~Myr in NGC~5253) by the harsh dynamic environment of the 
galaxies' centers.  

We compare the star clusters in these dwarf starburst galaxies to the 
clusters in the nuclear starburst of M~83, a giant spiral galaxy.  
The three cluster populations have similar age distributions, and 
they cover similar ranges in mass.  However, the clusters in M~83 
have a mass distribution that is more top-heavy.  Furthermore, the 
spatial distribution of clusters in M~83 is much more structured than 
in either dwarf.  Most of its clusters lie within a distinct annular 
ring structure centered on M~83's nucleus.  This contrast may result 
from a fundamental difference in the conditions that lead these 
galaxies to starburst.  M~83's strong bar feature is likely funneling 
material into the nuclear region, creating a localized region rich in 
gas and dust; an environment ripe for vigorous star formation.  No 
such internal dynamics mechanism exists in the dwarf galaxies, so 
something else must have caused them to starburst.  External 
interactions seem like a likely candidate for NGC~3077, given its 
well-known interactions within the M~81 group.  However, NGC~5253 may 
be too isolated to have been recently triggered by an interaction, so 
the cause for its current starburst activity is less clear.

\vskip 1in
\noindent Acknowledgements:
This work has been supported by the NASA LTSA grant NAG5-9173 and 
by the NASA HST grant GO-9144.01-A.

\clearpage

\clearpage

\begin{deluxetable}{lrrrrccr}
\tabletypesize{\scriptsize}
\tablecaption{Characteristics of Program Galaxies. \label{tab:galprop}}
\tablewidth{0pt}
\tablehead{
  \colhead{Galaxy} & \colhead{Morph.\tnm{a}} & \colhead{D}  & \colhead{M$_B$\tnm{a}} 
      & \colhead{angular size}\tnm{a} & \colhead{linear size} 
      & \colhead{$E(B-V)_{MW}$\tnm{b}} & \colhead{12$+\log$(O/H)}  \\
  \colhead{Name} & \colhead{Type} & \colhead{(Mpc)} & \colhead{(mag)} 
      & \colhead{(arcmin)} & \colhead{(kpc)} & \colhead{(mag)} & \colhead{} 
}

\startdata

NGC~3077        & I0 pec  & $3.2$\tnm{c}; $3.85\pm0.3$\tnm{d}        & $-17.5$ 
  & $5.4\times4.5$   & 5.6 & 0.07 & $8.9$\tnm{i} \\
NGC~5253        & Im pec  & $3.25\pm0.2$\tnm{e}; $3.9\pm0.5$\tnm{f}  & $-17.1$ 
  & $5.0\times1.9$   & 5.6 & 0.06 & $8.2$\tnm{i}; $8.23$\tnm{j} \\
NGC~5236 (M~83) & SAB(s)c & $3.7\pm0.2$\tnm{g}; $4.5\pm0.8$\tnm{h}   & $-20.0$ 
  & $12.9\times11.5$ &  16 & 0.07 & $9.16$\tnm{k} \\
 \enddata

\tablenotetext{a}{Morphological and photometric data from the RC3 \citep{rc3}.}
\tablenotetext{b}{\citet{sfd98}}
\tablenotetext{c}{\citet{tam68}}
\tablenotetext{d}{Tip of the Red Giant Branch (TRGB); \citet{kar02a}}
\tablenotetext{e}{Cepheid Variables; \citet{fre01}}
\tablenotetext{f}{TRGB; \citet{kar02b}}
\tablenotetext{g}{\citet{dev79}}
\tablenotetext{h}{Supernova Expanding Photosphere; \citet{sch94}}
\tablenotetext{i}{\citet{cal03}}
\tablenotetext{j}{\citet{mar97}}
\tablenotetext{k}{\citet{zar94}}

\end{deluxetable}

\begin{deluxetable}{rcclrc}
\tabletypesize{\scriptsize}
\tablecaption{Log of the Exposures \label{tab:observations}}
\tablewidth{0pt}
\tablehead{
  \colhead{Exposure} & \colhead{Target} & \colhead{Date} & \colhead{Filter} 
    & \colhead{$t_{exp}$} & \colhead{Position Angle} \\
  \colhead{ } & \colhead{Name} & \colhead{YYYY-MM-DD} & \colhead{ } 
    & \colhead{(sec)} & \colhead{(\arcdeg)}
}

\startdata
 u6eu0301 & NGC~3077 & 2001-05-22 & F300W ($UV$) &  800 & 143.88 \\
 u6eu0302 & NGC~3077 & 2001-05-22 & \nodata      &  800 & 143.88 \\
 u6eu0303 & NGC~3077 & 2001-05-22 & \nodata      &  800 & 143.88 \\
 u6eu0304 & NGC~3077 & 2001-05-22 & F547M ($V$)  &  600 & 143.88 \\
 u6eu0401 & NGC~3077 & 2001-05-23 & \nodata      &  600 & 143.37 \\
 u6eu0305 & NGC~3077 & 2001-05-22 & F814W ($I$)  &  300 & 143.88 \\
 u6eu0402 & NGC~3077 & 2001-05-23 & \nodata      &  400 & 143.37 \\
 u6eu0306 & NGC~3077 & 2001-05-22 & F487N ($H_\beta$)  &  700 & 143.88 \\
 u6eu0307 & NGC~3077 & 2001-05-22 & \nodata            &  700 & 143.88 \\
 u6eu0403 & NGC~3077 & 2001-05-23 & \nodata            & 1300 & 143.37 \\
 u6eu0406 & NGC~3077 & 2001-05-23 & \nodata            &  700 & 143.37 \\
 u6eu030c & NGC~3077 & 2001-05-23 & F656N ($H_\alpha$) &  300 & 143.88 \\
 u6eu030d & NGC~3077 & 2001-05-23 & \nodata            &  800 & 143.88 \\
 u6eu0405 & NGC~3077 & 2001-05-23 & \nodata            &  800 & 143.37 \\
\hline
u65r2602r & NGC~5253 & 2000-07-24 & F300W ($UV$) & 1000 & 156.40 \\
u65r2603r & NGC~5253 & 2000-07-24 & \nodata      &  800 & 156.40 \\
 u3760107 & NGC~5253 & 1996-05-08 & F547M ($V$)  &  200 & 129.34 \\
 u3760108 & NGC~5253 & 1996-05-08 & \nodata      &  600 & 129.34 \\
 u3760109 & NGC~5253 & 1996-05-08 & \nodata      &  200 & 129.34 \\
 u376010a & NGC~5253 & 1996-05-08 & \nodata      &  600 & 129.34 \\
 u3760105 & NGC~5253 & 1996-05-08 & F814W ($I$)  &  400 & 129.34 \\
 u3760106 & NGC~5253 & 1996-05-08 & \nodata      &  180 & 129.34 \\
 u376010b & NGC~5253 & 1996-05-08 & \nodata      &  180 & 129.34 \\
 u376010c & NGC~5253 & 1996-05-08 & \nodata      &  400 & 129.34 \\
 u3760101 & NGC~5253 & 1996-05-08 & F487N ($H_\beta$)  & 1200 & 129.34 \\
 u3760102 & NGC~5253 & 1996-05-08 & \nodata            & 1300 & 129.34 \\
 u376010f & NGC~5253 & 1996-05-09 & \nodata            & 1300 & 129.34 \\
 u376010g & NGC~5253 & 1996-05-09 & \nodata            & 1300 & 129.34 \\
 u3760103 & NGC~5253 & 1996-05-08 & F656N ($H_\alpha$) &  500 & 129.34 \\
 u3760104 & NGC~5253 & 1996-05-08 & \nodata            & 1500 & 129.34 \\
 u376010d & NGC~5253 & 1996-05-08 & \nodata            &  500 & 129.34 \\
 u376010e & NGC~5253 & 1996-05-08 & \nodata            & 1100 & 129.34 \\
\enddata
\end{deluxetable}

\clearpage
\thispagestyle{empty}
\begin{deluxetable}{rrrccccrrrrcccccc}
\rotate
\tabletypesize{\scriptsize}
\tablecolumns{14}
\tablewidth{0pt}
 
\tablecaption{Photometry of Clusters in NGC~3077 \label{tab:n3077clusters}}
\tablehead{
    \colhead{ID} & \colhead{$\alpha$(2000)} & \colhead{$\delta$(2000)} &
        \colhead{$m_{300}$} & \colhead{$m_{547}$} & \colhead{$m_{814}$} & \colhead{E(B-V)} & 
        \colhead{Phot. Mass\tnm{a}} & \colhead{Mass range} & \colhead{Phot. Age\tnm{a}} & \colhead{Age range\tnm{a}} & 
        \colhead{EW(H$\alpha$)} & \colhead{H$\alpha$\ Age\tnm{b}} & \colhead{Age range\tnm{b}} & 
        \colhead{Mass\tnm{c}} & \colhead{Age\tnm{c}} \\
 
    \colhead{ } & \colhead{ } & \colhead{ } & 
        \colhead{(STMAG)} & \colhead{(STMAG)} & \colhead{(STMAG)} & \colhead{(mag)} & 
        \colhead{($10^3 M_\odot$)} & \colhead{($10^3 M_\odot$)} & \colhead{Myr} & \colhead{Myr} & 
        \colhead{($\AA$)} & \colhead{(Myr)} & \colhead{(Myr)} & \colhead{($10^3 M_\odot$)} & \colhead{(Myr)} \\
}
 
\startdata
 1 & 10$^{h}$  3$^{m}$ 19.15$^{s}$ & 68\arcdeg\ 44\arcmin\  2.17\arcsec\ & 15.294 & 16.282 & 16.732 & 0.70 &     69 &  59-- 219 &      8 &   8--  14 &    2 &     18 &  14--  21 &  218 &   14 \\
 2 & 10$^{h}$  3$^{m}$ 17.42$^{s}$ & 68\arcdeg\ 43\arcmin\ 50.42\arcsec\ & 19.384 & 19.079 & 20.073 & 0.06 &     90 &  81-- 105 &    322 & 257-- 389 &    0 &\nodata &   \nodata &   90 &  322 \\
 3 & 10$^{h}$  3$^{m}$ 14.75$^{s}$ & 68\arcdeg\ 44\arcmin\  0.17\arcsec\ & 18.460 & 18.810 & 19.624 & 0.06 &     70 &  57--  84 &    113 &  65-- 170 &    0 &\nodata &   \nodata &   70 &  113 \\
 4 & 10$^{h}$  3$^{m}$ 17.64$^{s}$ & 68\arcdeg\ 43\arcmin\ 56.21\arcsec\ & 17.565 & 18.289 & 19.184 & 0.28 &     67 &   7--  77 &     70 &   7--  77 &    0 &\nodata &   \nodata &   67 &   70 \\
 5 & 10$^{h}$  3$^{m}$ 18.14$^{s}$ & 68\arcdeg\ 43\arcmin\ 59.75\arcsec\ & 19.585 & 19.789 & 20.391 & 0.06 &     33 &  24--  50 &     68 &  54-- 187 &    0 &\nodata &   \nodata &   33 &   68 \\
 6 & 10$^{h}$  3$^{m}$ 18.69$^{s}$ & 68\arcdeg\ 43\arcmin\ 59.47\arcsec\ & 16.145 & 17.467 & 18.769 & 1.65 &     22 &  12-- 104 &      6 &   1--  33 &  217 &      6 &   6--   6 &   22 &    6 \\
 7 & 10$^{h}$  3$^{m}$ 20.09$^{s}$ & 68\arcdeg\ 44\arcmin\  6.30\arcsec\ & 15.943 & 17.700 & 19.034 & 0.06 &     16 &  11--  20 &      6 &   4--   6 &   44 &      7 &   7--   7 &   20 &    6 \\
 8 & 10$^{h}$  3$^{m}$ 17.29$^{s}$ & 68\arcdeg\ 43\arcmin\ 54.12\arcsec\ & 20.036 & 20.314 & 21.291 & 0.06 &     19 &   8--  33 &    170 &  37-- 264 &    0 &\nodata &   \nodata &   19 &  170 \\
 9 & 10$^{h}$  3$^{m}$ 18.11$^{s}$ & 68\arcdeg\ 43\arcmin\ 48.06\arcsec\ & 20.106 & 20.390 & 21.382 & 0.06 &     18 &  10--  28 &    170 &  52-- 250 &    0 &\nodata &   \nodata &   18 &  170 \\
10 & 10$^{h}$  3$^{m}$ 22.69$^{s}$ & 68\arcdeg\ 44\arcmin\  1.39\arcsec\ & 20.182 & 20.502 & 21.001 & 0.06 &     16 &  15--  18 &     68 &  65--  90 &    0 &\nodata &   \nodata &   16 &   68 \\
11 & 10$^{h}$  3$^{m}$ 18.06$^{s}$ & 68\arcdeg\ 43\arcmin\ 55.67\arcsec\ & 16.898 & 17.974 & 18.385 & 0.70 &     16 &  13--  48 &      8 &   8--  13 &    0 &\nodata &   \nodata &   16 &    8 \\
12 & 10$^{h}$  3$^{m}$ 17.02$^{s}$ & 68\arcdeg\ 44\arcmin\  3.29\arcsec\ & 19.968 & 20.408 & 21.120 & 0.21 &     13 &   8--  21 &     50 &  36-- 157 &    0 &\nodata &   \nodata &   13 &   50 \\
13 & 10$^{h}$  3$^{m}$ 17.45$^{s}$ & 68\arcdeg\ 43\arcmin\ 55.35\arcsec\ & 20.750 & 20.981 & 22.072 & 0.06 &     11 &   0--  22 &    183 &   7-- 378 &    0 &\nodata &   \nodata &   11 &  183 \\
14 & 10$^{h}$  3$^{m}$ 18.91$^{s}$ & 68\arcdeg\ 44\arcmin\  5.56\arcsec\ & 17.227 & 18.326 & 19.565 & 0.69\tnm{d} &      9 &   6--  80 &      6 &   3--  60 &   93 &      6 &   4--   9 &    9 &    6 \\
15 & 10$^{h}$  3$^{m}$ 19.71$^{s}$ & 68\arcdeg\ 44\arcmin\  6.03\arcsec\ & 18.546 & 19.446 & 20.410 & 0.36 &      9 &   2--  54 &     16 &   6-- 139 &    0 &\nodata &   \nodata &    9 &   16 \\
16 & 10$^{h}$  3$^{m}$ 19.62$^{s}$ & 68\arcdeg\ 44\arcmin\  3.88\arcsec\ & 18.713 & 19.473 & 20.385 & 0.44 &      8 &   2-- 109 &     15 &   4-- 286 &    0 &\nodata &   \nodata &    8 &   15 \\
17 & 10$^{h}$  3$^{m}$ 19.86$^{s}$ & 68\arcdeg\ 44\arcmin\  5.46\arcsec\ & 17.128 & 19.410 & 20.439 & 0.51 &      4 &   3--  15 &      1 &   1--  12 &  518 &      4 &   3--   6 &    8 &    4 \\
18 & 10$^{h}$  3$^{m}$ 17.72$^{s}$ & 68\arcdeg\ 43\arcmin\ 49.54\arcsec\ & 20.368 & 20.925 & 21.726 & 0.06 &      7 &   1--  15 &     73 &   7-- 191 &    0 &\nodata &   \nodata &    8 &   73 \\
19 & 10$^{h}$  3$^{m}$ 19.00$^{s}$ & 68\arcdeg\ 44\arcmin\  4.14\arcsec\ & 17.265 & 18.917 & 20.037 & 0.35\tnm{d} &      7 &   4--   7 &      5 &   3--   6 &  442 &      5 &   4--   6 &    7 &    5 \\
20 & 10$^{h}$  3$^{m}$ 18.33$^{s}$ & 68\arcdeg\ 43\arcmin\ 59.85\arcsec\ & 17.604 & 18.999 & 19.813 & 0.81 &     10 &   4--  31 &     14 &   4--  36 &   24 &      7 &   6--  13 &    7 &    7 \\
21 & 10$^{h}$  3$^{m}$ 19.60$^{s}$ & 68\arcdeg\ 44\arcmin\  3.47\arcsec\ & 18.435 & 19.505 & 20.265 & 0.43 &      7 &   2--  74 &     14 &   4-- 174 &    0 &\nodata &   \nodata &    7 &   14 \\
22 & 10$^{h}$  3$^{m}$ 19.93$^{s}$ & 68\arcdeg\ 44\arcmin\  2.97\arcsec\ & 17.681 & 19.031 & 20.257 & 0.63\tnm{d} &      6 &   3--  60 &      6 &   1--  83 &  360 &      6 &   4--   6 &    6 &    6 \\
23 & 10$^{h}$  3$^{m}$ 19.12$^{s}$ & 68\arcdeg\ 44\arcmin\  6.04\arcsec\ & 17.625 & 18.901 & 20.157 & 0.25\tnm{d} &      6 &   5--  17 &      6 &   6--  18 &   32 &      7 &   6--  12 &    6 &    6 \\
24 & 10$^{h}$  3$^{m}$ 19.42$^{s}$ & 68\arcdeg\ 44\arcmin\  4.58\arcsec\ & 17.315 & 18.646 & 19.741 & 0.49 &      6 &   6--  26 &      6 &   5--  22 &  400 &      5 &   4--   6 &    6 &    6 \\
25 & 10$^{h}$  3$^{m}$ 19.44$^{s}$ & 68\arcdeg\ 44\arcmin\  5.24\arcsec\ & 17.141 & 18.496 & 19.907 & 0.60\tnm{d} &      8 &   5--  26 &      6 &   4--  20 &  555 &      4 &   4--   6 &    6 &    4 \\
26 & 10$^{h}$  3$^{m}$ 18.27$^{s}$ & 68\arcdeg\ 44\arcmin\  1.72\arcsec\ & 17.657 & 18.687 & 19.906 & 0.61\tnm{d} &     13 &   5--  24 &     16 &   6--  25 &   47 &      7 &   6--   7 &    6 &    7 \\
27 & 10$^{h}$  3$^{m}$ 20.23$^{s}$ & 68\arcdeg\ 44\arcmin\  7.17\arcsec\ & 17.590 & 19.100 & 20.564 & 0.06 &      4 &   4--   5 &      6 &   6--   6 &   74 &      6 &   6--   7 &    5 &    6 \\
28 & 10$^{h}$  3$^{m}$ 19.96$^{s}$ & 68\arcdeg\ 44\arcmin\  8.98\arcsec\ & 17.534 & 18.491 & 19.436 & 0.27\tnm{d} &     29 &   5--  42 &     23 &   7--  39 &   38 &      7 &   7--   7 &    5 &    7 \\
29 & 10$^{h}$  3$^{m}$ 17.91$^{s}$ & 68\arcdeg\ 43\arcmin\ 47.38\arcsec\ & 19.741 & 20.309 & 20.965 & 0.06 &      5 &   1--  23 &     15 &   7-- 152 &    0 &\nodata &   \nodata &    5 &   15 \\
30 & 10$^{h}$  3$^{m}$ 18.36$^{s}$ & 68\arcdeg\ 43\arcmin\ 58.10\arcsec\ & 17.792 & 18.569 & 20.014 & 0.80 &     14 &   5--  49 &     16 &   6--  56 &  140 &      6 &   6--   6 &    5 &    6 \\
31 & 10$^{h}$  3$^{m}$ 20.29$^{s}$ & 68\arcdeg\ 44\arcmin\  7.71\arcsec\ & 17.732 & 19.429 & 20.753 & 0.06\tnm{d} &      3 &   2--   4 &      6 &   4--   6 &   46 &      7 &   6--   7 &    5 &    6 \\
32 & 10$^{h}$  3$^{m}$ 22.58$^{s}$ & 68\arcdeg\ 44\arcmin\  1.81\arcsec\ & 18.864 & 20.042 & 20.844 & 0.06 &      4 &   4--   4 &     14 &  14--  14 &    0 &\nodata &   \nodata &    4 &   14 \\
33 & 10$^{h}$  3$^{m}$ 19.48$^{s}$ & 68\arcdeg\ 44\arcmin\  5.92\arcsec\ & 18.010 & 19.448 & 20.653 & 0.37\tnm{d} &      4 &   2--  14 &      6 &   3--  24 &  170 &      6 &   6--   6 &    4 &    6 \\
34 & 10$^{h}$  3$^{m}$ 19.34$^{s}$ & 68\arcdeg\ 44\arcmin\  5.50\arcsec\ & 18.034 & 19.705 & 20.408 & 0.22 &      9 &   6--  10 &     12 &   9--  12 &  192 &      6 &   6--   6 &    4 &    6 \\
35 & 10$^{h}$  3$^{m}$ 19.13$^{s}$ & 68\arcdeg\ 44\arcmin\  3.73\arcsec\ & 17.790 & 19.302 & 20.544 & 0.51 &      4 &   2--  24 &      6 &   1--  36 &  613 &      4 &   2--   6 &    3 &    4 \\
36 & 10$^{h}$  3$^{m}$ 19.21$^{s}$ & 68\arcdeg\ 43\arcmin\ 58.70\arcsec\ & 17.505 & 19.701 & 20.778 & 0.56 &      3 &   2--   4 &      1 &   1--   5 &  911 &      3 &   0--   6 &    3 &    1 \\
37 & 10$^{h}$  3$^{m}$ 18.90$^{s}$ & 68\arcdeg\ 44\arcmin\  6.72\arcsec\ & 18.549 & 19.914 & 21.551 & 0.18 &      2 &   1--   9 &      6 &   4--  26 &   48 &      7 &   4--  13 &    2 &    6 \\
38 & 10$^{h}$  3$^{m}$ 18.92$^{s}$ & 68\arcdeg\ 43\arcmin\ 54.99\arcsec\ & 18.432 & 19.414 & 20.566 & 0.56 &      7 &   2--  30 &     16 &   6--  70 &  124 &      6 &   6--   6 &    2 &    6 \\
39 & 10$^{h}$  3$^{m}$ 19.77$^{s}$ & 68\arcdeg\ 44\arcmin\  6.72\arcsec\ & 18.827 & 19.807 & 21.323 & 0.15 &      5 &   2--  10 &     16 &   6--  31 &   42 &      7 &   6--   9 &    2 &    7 \\
40 & 10$^{h}$  3$^{m}$ 18.39$^{s}$ & 68\arcdeg\ 43\arcmin\ 54.08\arcsec\ & 19.772 & 20.388 & 21.166 & 0.06 &      9 &   1--  22 &     44 &   7-- 174 &   18 &      8 &   7--  13 &    2 &    8 \\
41 & 10$^{h}$  3$^{m}$ 19.28$^{s}$ & 68\arcdeg\ 43\arcmin\ 51.77\arcsec\ & 19.681 & 20.249 & 21.315 & 0.26\tnm{d} &     14 &   1--  32 &    165 &   6-- 236 &   19 &      8 &   7--  13 &    2 &    8 \\
42 & 10$^{h}$  3$^{m}$ 18.98$^{s}$ & 68\arcdeg\ 44\arcmin\  6.46\arcsec\ & 19.128 & 19.920 & 20.756 & 0.11 &     11 &   1--  35 &     39 &   6-- 143 &   69 &      6 &   6--   7 &    2 &    6 \\
43 & 10$^{h}$  3$^{m}$ 19.26$^{s}$ & 68\arcdeg\ 44\arcmin\  6.75\arcsec\ & 19.467 & 19.805 & 21.087 & 0.06 &     27 &   1--  56 &    161 &   7-- 300 &   64 &      7 &   5--  12 &    1 &    7 \\
44 & 10$^{h}$  3$^{m}$ 18.54$^{s}$ & 68\arcdeg\ 44\arcmin\ 10.44\arcsec\ & 19.856 & 20.591 & 21.343 & 0.31 &      1 &   1--  40 &      7 &   4-- 293 &    8 &     12 &   8--  14 &    1 &    8 \\
45 & 10$^{h}$  3$^{m}$ 20.21$^{s}$ & 68\arcdeg\ 44\arcmin\  5.42\arcsec\ & 18.243 & 20.125 & 21.805 & 0.21 &      1 &   1--   2 &      4 &   1--   6 &    0 &\nodata &   \nodata &    1 &    4 \\
46 & 10$^{h}$  3$^{m}$ 18.37$^{s}$ & 68\arcdeg\ 44\arcmin\  5.66\arcsec\ & 20.087 & 20.548 & 21.045 & 0.15 &     14 &   1--  43 &     80 &   7-- 322 &   92 &      6 &   6--   7 &    1 &    7 \\
47 & 10$^{h}$  3$^{m}$ 20.97$^{s}$ & 68\arcdeg\ 44\arcmin\  8.40\arcsec\ & 19.936 & 20.858 & 21.655 & 0.06 &      1 &   1--   7 &      7 &   6--  68 &   13 &     11 &   7--  13 &    1 &    7 \\
48 & 10$^{h}$  3$^{m}$ 17.30$^{s}$ & 68\arcdeg\ 44\arcmin\  8.28\arcsec\ & 19.753 & 20.945 & 21.932 & 0.12 &      1 &   1--   7 &      7 &   4--  65 &    0 &\nodata &   \nodata &    1 &    7 \\
49 & 10$^{h}$  3$^{m}$ 19.22$^{s}$ & 68\arcdeg\ 44\arcmin\  4.31\arcsec\ & 19.035 & 20.259 & 99.999 & 0.15 &\nodata &   \nodata &\nodata &   \nodata &  123 &      6 &   6--   7 & \nodata &    6 \\
50 & 10$^{h}$  3$^{m}$ 18.91$^{s}$ & 68\arcdeg\ 43\arcmin\ 58.99\arcsec\ & 99.999 & 19.174 & 20.036 & 0.35 &\nodata &   \nodata &\nodata &   \nodata &   78 &      6 &   6--   7 & \nodata &    6 \\
51 & 10$^{h}$  3$^{m}$ 18.28$^{s}$ & 68\arcdeg\ 44\arcmin\  2.24\arcsec\ & 99.999 & 19.372 & 20.395 & 0.42 &\nodata &   \nodata &\nodata &   \nodata &   65 &      7 &   6--   7 & \nodata &    7 \\
52 & 10$^{h}$  3$^{m}$ 18.03$^{s}$ & 68\arcdeg\ 43\arcmin\ 58.78\arcsec\ & 99.999 & 20.041 & 20.839 & 0.41 &\nodata &   \nodata &\nodata &   \nodata &   12 &     11 &   7--  15 & \nodata &   11 \\
53 & 10$^{h}$  3$^{m}$ 21.02$^{s}$ & 68\arcdeg\ 44\arcmin\  1.76\arcsec\ & 18.733 & 19.310 & 20.829 & 0.46 &\nodata &   \nodata &\nodata &   \nodata &   11 &     12 &   7--  15 & \nodata &   12 \\
54 & 10$^{h}$  3$^{m}$ 17.80$^{s}$ & 68\arcdeg\ 43\arcmin\ 59.93\arcsec\ & 99.999 & 20.493 & 21.454 & 0.06 &\nodata &   \nodata &\nodata &   \nodata &    1 &     21 &   5--  30 & \nodata &   21 \\
55 & 10$^{h}$  3$^{m}$ 18.22$^{s}$ & 68\arcdeg\ 44\arcmin\  0.06\arcsec\ & 99.999 & 19.649 & 19.445 & 0.06 &\nodata &   \nodata &\nodata &   \nodata &    0 &\nodata &   \nodata & \nodata & \nodata \\
\enddata
\tablenotetext{a}{Mass and Age estimates from broad-band photometry.  There are no 
estimates for points which are farther than 5$\sigma$ from the Starburst99 model track.}
\tablenotetext{b}{Age estimates from EW(H$\alpha$).  There is no estimate when 
EW(H$\alpha$) is not detected, although these clusters are likely older than 10--20~Myr.}
\tablenotetext{c}{Final Mass and Age estimates, combining constraints from 
the broad-band photometry and EW(H$\alpha$).}
\tablenotetext{d}{The E(B-V) estimate for this object was manually adjusted to bring
its photometric age estimate into agreement with its EW(H$\alpha$).}
\end{deluxetable}

\clearpage
\thispagestyle{empty}
\begin{deluxetable}{rrrccccrrrrcccccc}
\rotate
\tabletypesize{\scriptsize}
\tablecolumns{14}
\tablewidth{0pt}
 
\tablecaption{Photometry of Clusters in NGC~5253 \label{tab:n5253clusters}}
\tablehead{
    \colhead{ID} & \colhead{$\alpha$(2000)} & \colhead{$\delta$(2000)} &
        \colhead{$m_{300}$} & \colhead{$m_{547}$} & \colhead{$m_{814}$} & \colhead{E(B-V)} & 
        \colhead{Phot. Mass\tnm{a}} & \colhead{Mass range} & \colhead{Phot. Age\tnm{a}} & \colhead{Age range\tnm{a}} & 
        \colhead{EW(H$\alpha$)} & \colhead{H$\alpha$\ Age\tnm{b}} & \colhead{Age range\tnm{b}} & 
        \colhead{Mass\tnm{c}} & \colhead{Age\tnm{c}} \\
 
    \colhead{ } & \colhead{ } & \colhead{ } & 
        \colhead{(STMAG)} & \colhead{(STMAG)} & \colhead{(STMAG)} & \colhead{(mag)} & 
        \colhead{($10^3 M_\odot$)} & \colhead{($10^3 M_\odot$)} & \colhead{Myr} & \colhead{Myr} & 
        \colhead{($\AA$)} & \colhead{(Myr)} & \colhead{(Myr)} & \colhead{($10^3 M_\odot$)} & \colhead{(Myr)} \\
}
 
\startdata
 1 & 13$^{h}$ 39$^{m}$ 56.01$^{s}$ & -31\arcdeg\ 38\arcmin\ 25.05\arcsec\ & 13.940 & 16.054 & 17.015 & 0.96\tnm{d} &    103 &  86-- 120 &      1 &   1--   1 & 1981 &      3 &   3--   3 &  118 &    3 \\
 2 & 13$^{h}$ 39$^{m}$ 55.98$^{s}$ & -31\arcdeg\ 38\arcmin\ 31.87\arcsec\ & 15.800 & 17.152 & 17.641 & 0.06 &     46 &  42--  52 &     10 &  10--  11 &    0 &\nodata &   \nodata &   46 &   10 \\
 3 & 13$^{h}$ 39$^{m}$ 55.53$^{s}$ & -31\arcdeg\ 38\arcmin\ 29.75\arcsec\ & 16.756 & 17.799 & 17.866 & 0.06 &     58 &  44--  70 &     12 &  11--  14 &   14 &     11 &  10--  12 &   42 &   11 \\
 4 & 13$^{h}$ 39$^{m}$ 55.93$^{s}$ & -31\arcdeg\ 38\arcmin\ 27.48\arcsec\ & 15.301 & 17.424 & 18.359 & 0.13 &     27 &  26--  27 &      1 &   1--   1 &  505 &      5 &   5--   5 &   27 &    1 \\
 5 & 13$^{h}$ 39$^{m}$ 55.58$^{s}$ & -31\arcdeg\ 38\arcmin\ 29.36\arcsec\ & 16.551 & 17.431 & 18.029 & 0.06 &     22 &  18--  59 &      8 &   8--  15 &   17 &     10 &   9--  11 &   21 &    8 \\
 6 & 13$^{h}$ 39$^{m}$ 55.87$^{s}$ & -31\arcdeg\ 38\arcmin\ 26.87\arcsec\ & 15.718 & 17.755 & 18.898 & 0.70\tnm{d} &     16 &  13--  83 &      2 &   1--   6 &  884 &      4 &   4--   5 &   20 &    4 \\
 7 & 13$^{h}$ 39$^{m}$ 56.24$^{s}$ & -31\arcdeg\ 38\arcmin\ 28.33\arcsec\ & 15.771 & 17.267 & 18.676 & 0.65\tnm{d} &     17 &  17--  19 &      6 &   4--   6 &  438 &      5 &   5--   5 &   18 &    5 \\
 8 & 13$^{h}$ 39$^{m}$ 55.98$^{s}$ & -31\arcdeg\ 38\arcmin\ 27.99\arcsec\ & 16.681 & 18.470 & 19.503 & 0.06 &     12 &   6--  12 &      3 &   3--   5 &  324 &      5 &   5--   5 &   13 &    5 \\
 9 & 13$^{h}$ 39$^{m}$ 55.37$^{s}$ & -31\arcdeg\ 38\arcmin\ 33.96\arcsec\ & 16.667 & 17.928 & 18.458 & 0.06 &     20 &  18--  32 &     10 &   9--  14 &   34 &      8 &   8--   9 &   12 &    8 \\
10 & 13$^{h}$ 39$^{m}$ 56.08$^{s}$ & -31\arcdeg\ 38\arcmin\ 31.77\arcsec\ & 17.887 & 19.164 & 19.579 & 0.06 &      8 &   5--  16 &     10 &   9--  15 &    3 &     16 &  13--  22 &    8 &   10 \\
11 & 13$^{h}$ 39$^{m}$ 55.51$^{s}$ & -31\arcdeg\ 38\arcmin\ 24.52\arcsec\ & 18.758 & 20.033 & 19.888 & 0.06 &      9 &   5--  10 &     13 &  12--  14 &   61 &      7 &   6--  11 &    5 &   11 \\
12 & 13$^{h}$ 39$^{m}$ 55.87$^{s}$ & -31\arcdeg\ 38\arcmin\ 33.51\arcsec\ & 17.599 & 19.652 & 20.495 & 0.06 &      4 &   3--   5 &      1 &   1--   4 &    6 &     14 &  11--  20 &    4 &    1 \\
13 & 13$^{h}$ 39$^{m}$ 55.94$^{s}$ & -31\arcdeg\ 38\arcmin\ 22.07\arcsec\ & 18.124 & 19.705 & 20.390 & 0.06 &      4 &   2--  14 &      4 &   4--  28 &  554 &      5 &   5--   5 &    4 &    4 \\
14 & 13$^{h}$ 39$^{m}$ 55.45$^{s}$ & -31\arcdeg\ 38\arcmin\ 33.69\arcsec\ & 19.382 & 20.598 & 20.258 & 0.06 &      6 &   4--   6 &     13 &  12--  13 &  242 &      6 &   5--   6 &    4 &   12 \\
15 & 13$^{h}$ 39$^{m}$ 55.61$^{s}$ & -31\arcdeg\ 38\arcmin\ 32.41\arcsec\ & 18.407 & 20.326 & 21.319 & 0.06 &      2 &   1--   6 &      3 &   1--  19 &  410 &      5 &   5--   6 &    4 &    5 \\
16 & 13$^{h}$ 39$^{m}$ 57.58$^{s}$ & -31\arcdeg\ 38\arcmin\ 19.25\arcsec\ & 18.526 & 19.681 & 20.073 & 0.14 &      5 &   3--   8 &     10 &   9--  14 &   36 &      8 &   8--   8 &    3 &    8 \\
17 & 13$^{h}$ 39$^{m}$ 55.76$^{s}$ & -31\arcdeg\ 38\arcmin\ 31.69\arcsec\ & 17.877 & 19.924 & 20.747 & 0.11 &      3 &   2--   3 &      1 &   1--   4 & 1056 &      4 &   3--   5 &    3 &    4 \\
18 & 13$^{h}$ 39$^{m}$ 55.49$^{s}$ & -31\arcdeg\ 38\arcmin\ 25.40\arcsec\ & 18.007 & 20.069 & 21.923 & 0.06 &      1 &   1--   2 &      4 &   1--   5 &  296 &      5 &   5--   6 &    2 &    5 \\
19 & 13$^{h}$ 39$^{m}$ 56.57$^{s}$ & -31\arcdeg\ 38\arcmin\ 22.82\arcsec\ & 18.156 & 19.556 & 20.758 & 0.06 &      2 &   2--  12 &      5 &   5--  24 &  234 &      6 &   5--   6 &    2 &    6 \\
20 & 13$^{h}$ 39$^{m}$ 55.92$^{s}$ & -31\arcdeg\ 38\arcmin\ 24.56\arcsec\ & 17.807 & 19.404 & 99.999 & 0.27 &\nodata &   \nodata &\nodata &   \nodata & 1820 &      3 &   3--   3 & \nodata &    3 \\
21 & 13$^{h}$ 39$^{m}$ 55.83$^{s}$ & -31\arcdeg\ 38\arcmin\ 27.05\arcsec\ & 17.048 & 18.593 & 19.371 & 0.13 &\nodata &   \nodata &\nodata &   \nodata &  920 &      4 &   3--   5 & \nodata &    4 \\
22 & 13$^{h}$ 39$^{m}$ 56.04$^{s}$ & -31\arcdeg\ 38\arcmin\ 38.06\arcsec\ & 18.215 & 19.917 & 19.764 & 0.13 &\nodata &   \nodata &\nodata &   \nodata &  970 &      4 &   4--   4 & \nodata &    4 \\
23 & 13$^{h}$ 39$^{m}$ 56.26$^{s}$ & -31\arcdeg\ 38\arcmin\ 26.18\arcsec\ & 99.999 & 19.008 & 19.225 & 0.06 &\nodata &   \nodata &\nodata &   \nodata &  682 &      5 &   4--   5 & \nodata &    5 \\
24 & 13$^{h}$ 39$^{m}$ 55.92$^{s}$ & -31\arcdeg\ 38\arcmin\ 27.00\arcsec\ & 16.581 & 18.971 & 19.609 & 0.19 &\nodata &   \nodata &\nodata &   \nodata &  463 &      5 &   5--   5 & \nodata &    5 \\
25 & 13$^{h}$ 39$^{m}$ 55.94$^{s}$ & -31\arcdeg\ 38\arcmin\ 28.18\arcsec\ & 17.116 & 19.077 & 19.562 & 0.06 &\nodata &   \nodata &\nodata &   \nodata &  454 &      5 &   5--   5 & \nodata &    5 \\
26 & 13$^{h}$ 39$^{m}$ 55.73$^{s}$ & -31\arcdeg\ 38\arcmin\ 30.89\arcsec\ & 17.499 & 19.472 & 99.999 & 0.06 &\nodata &   \nodata &\nodata &   \nodata &  623 &      5 &   3--   6 & \nodata &    5 \\
27 & 13$^{h}$ 39$^{m}$ 55.86$^{s}$ & -31\arcdeg\ 38\arcmin\ 38.90\arcsec\ & 16.895 & 18.515 & 18.927 & 0.06 &\nodata &   \nodata &\nodata &   \nodata &  143 &      6 &   6--   6 & \nodata &    6 \\
28 & 13$^{h}$ 39$^{m}$ 55.72$^{s}$ & -31\arcdeg\ 38\arcmin\ 38.26\arcsec\ & 19.209 & 19.368 & 20.490 & 0.06 &\nodata &   \nodata &\nodata &   \nodata &  105 &      6 &   6--   7 & \nodata &    6 \\
29 & 13$^{h}$ 39$^{m}$ 55.44$^{s}$ & -31\arcdeg\ 38\arcmin\ 29.62\arcsec\ & 18.492 & 20.309 & 20.025 & 0.06 &\nodata &   \nodata &\nodata &   \nodata &   65 &      7 &   7--   7 & \nodata &    7 \\
30 & 13$^{h}$ 39$^{m}$ 56.70$^{s}$ & -31\arcdeg\ 38\arcmin\ 19.02\arcsec\ & 18.943 & 19.766 & 99.999 & 0.06 &\nodata &   \nodata &\nodata &   \nodata &   37 &      8 &   7--   9 & \nodata &    8 \\
31 & 13$^{h}$ 39$^{m}$ 55.73$^{s}$ & -31\arcdeg\ 38\arcmin\ 16.78\arcsec\ & 18.505 & 19.112 & 19.286 & 0.14 &\nodata &   \nodata &\nodata &   \nodata &    0 &\nodata &   \nodata & \nodata & \nodata \\
32 & 13$^{h}$ 39$^{m}$ 56.01$^{s}$ & -31\arcdeg\ 38\arcmin\ 22.41\arcsec\ & 18.381 & 19.118 & 19.233 & 0.13 &\nodata &   \nodata &\nodata &   \nodata &    0 &\nodata &   \nodata & \nodata & \nodata \\
33 & 13$^{h}$ 39$^{m}$ 55.38$^{s}$ & -31\arcdeg\ 38\arcmin\ 44.11\arcsec\ & 19.166 & 19.896 & 21.239 & 0.06 &\nodata &   \nodata &\nodata &   \nodata &    0 &\nodata &   \nodata & \nodata & \nodata \\
\enddata
\tablenotetext{a}{Mass and Age estimates from broad-band photometry.  There are no 
estimates for points which are farther than 5$\sigma$ from the Starburst99 model track.}
\tablenotetext{b}{Age estimates from EW(H$\alpha$).  There is no estimate when 
EW(H$\alpha$) is not detected, although these clusters are likely older than 10--20~Myr.}
\tablenotetext{c}{Final Mass and Age estimates, combining constraints from 
the broad-band photometry and EW(H$\alpha$).}
\tablenotetext{d}{The E(B-V) estimate for this object was manually adjusted to bring
its photometric age estimate into agreement with its EW(H$\alpha$).}
\end{deluxetable}

\clearpage
\begin{deluxetable}{rrrccccc}
\tabletypesize{\scriptsize}
\tablecolumns{8}
\tablewidth{0pt}
 
\tablecaption{Photometry of Non-cluster Objects in NGC~3077 \label{tab:n3077.extra}}
\tablehead{
    \colhead{ID} & \colhead{$\alpha$(2000)} & \colhead{$\delta$(2000)} &
        \colhead{$m_{300}$} & \colhead{$m_{547}$} & \colhead{$m_{814}$} & \colhead{E(B-V)} & \colhead{Type}\tnm{a} \\
 
    \colhead{ } & \colhead{ } & \colhead{ } & 
        \colhead{(STMAG)} & \colhead{(STMAG)} & \colhead{(STMAG)} & \colhead{(mag)} & \colhead{ } \\
}
 
\startdata
1 & 10$^{h}$ 03$^{m}$ 22.74$^{s}$ & 68\arcdeg\ 44\arcmin\ 27.26\arcsec\ &  21.354 &  17.954 &  17.596 &  0.06 & 1 \\
2 & 10$^{h}$ 03$^{m}$ 18.25$^{s}$ & 68\arcdeg\ 43\arcmin\ 59.75\arcsec\ &  18.069 &  19.030 &  19.825 &  0.81 & 2 \\
3 & 10$^{h}$ 03$^{m}$ 19.12$^{s}$ & 68\arcdeg\ 43\arcmin\ 55.41\arcsec\ &  18.797 &  19.346 &  20.129 &  0.27 & 1 \\
4 & 10$^{h}$ 03$^{m}$ 21.37$^{s}$ & 68\arcdeg\ 44\arcmin\ 17.47\arcsec\ &  18.553 &  19.940 &  21.011 &  0.06 & 1 \\
5 & 10$^{h}$ 03$^{m}$ 19.64$^{s}$ & 68\arcdeg\ 43\arcmin\ 55.69\arcsec\ &  18.972 &  20.001 &  21.435 &  0.22 & 1 \\
6 & 10$^{h}$ 03$^{m}$ 20.37$^{s}$ & 68\arcdeg\ 44\arcmin\  6.16\arcsec\ &  18.310 &  20.573 &  22.397 &  0.47 & 2 \\
7 & 10$^{h}$ 03$^{m}$ 20.55$^{s}$ & 68\arcdeg\ 44\arcmin\  4.81\arcsec\ &  20.285 &  20.769 &  21.483 &  0.06 & 1 \\
8 & 10$^{h}$ 03$^{m}$ 18.61$^{s}$ & 68\arcdeg\ 44\arcmin\  7.49\arcsec\ &  20.663 &  20.970 &  21.677 &  0.06 & 1 \\
9 & 10$^{h}$ 03$^{m}$ 17.58$^{s}$ & 68\arcdeg\ 43\arcmin\ 57.11\arcsec\ &  20.604 &  20.970 &  24.160 &  0.20 & 2 \\
10 & 10$^{h}$ 03$^{m}$ 19.05$^{s}$ & 68\arcdeg\ 44\arcmin\  8.98\arcsec\ &  20.513 &  21.117 &  22.394 &  0.18 & 1 \\
\enddata
\tablenotetext{a}{1=star; 2=questionable star/cluster classification; 3=detected in only one filter}
\end{deluxetable}

\begin{deluxetable}{rrrccccc}
\tabletypesize{\scriptsize}
\tablecolumns{8}
\tablewidth{0pt}
 
\tablecaption{Photometry of Non-cluster Objects in NGC~5253 \label{tab:n5253.extra}}
\tablehead{
    \colhead{ID} & \colhead{$\alpha$(2000)} & \colhead{$\delta$(2000)} &
        \colhead{$m_{300}$} & \colhead{$m_{547}$} & \colhead{$m_{814}$} & \colhead{E(B-V)} & \colhead{Type}\tnm{a} \\
 
    \colhead{ } & \colhead{ } & \colhead{ } & 
        \colhead{(STMAG)} & \colhead{(STMAG)} & \colhead{(STMAG)} & \colhead{(mag)} & \colhead{ } \\
}
 
\startdata
1 & 13$^{h}$ 39$^{m}$ 57.59$^{s}$ & -31\arcdeg\ 38\arcmin\ 12.97\arcsec\ &  18.766 &  17.760 &  18.033 &  0.06 & 1 \\
2 & 13$^{h}$ 39$^{m}$ 54.35$^{s}$ & -31\arcdeg\ 38\arcmin\ 34.19\arcsec\ &  18.424 &  17.781 &  18.330 &  0.06 & 1 \\
3 & 13$^{h}$ 39$^{m}$ 55.88$^{s}$ & -31\arcdeg\ 38\arcmin\ 27.78\arcsec\ &  17.099 &  18.329 &  18.799 &  0.12 & 2 \\
4 & 13$^{h}$ 39$^{m}$ 55.96$^{s}$ & -31\arcdeg\ 38\arcmin\ 38.62\arcsec\ &  17.653 &  18.569 &  19.446 &  0.06 & 1 \\
5 & 13$^{h}$ 39$^{m}$ 55.68$^{s}$ & -31\arcdeg\ 38\arcmin\ 28.18\arcsec\ &  18.880 &  18.751 &  19.922 &  0.06 & 2 \\
6 & 13$^{h}$ 39$^{m}$ 55.43$^{s}$ & -31\arcdeg\ 38\arcmin\ 27.78\arcsec\ &  19.217 &  18.778 &  19.702 &  0.06 & 1 \\
7 & 13$^{h}$ 39$^{m}$ 55.81$^{s}$ & -31\arcdeg\ 38\arcmin\ 34.37\arcsec\ &  17.940 &  18.821 &  19.765 &  0.11 & 2 \\
8 & 13$^{h}$ 39$^{m}$ 55.67$^{s}$ & -31\arcdeg\ 38\arcmin\ 37.61\arcsec\ &  17.270 &  18.836 &  19.672 &  0.06 & 1 \\
9 & 13$^{h}$ 39$^{m}$ 57.38$^{s}$ & -31\arcdeg\ 38\arcmin\ 22.72\arcsec\ &  19.178 &  19.077 &  19.895 &  0.06 & 1 \\
10 & 13$^{h}$ 39$^{m}$ 55.99$^{s}$ & -31\arcdeg\ 38\arcmin\ 11.59\arcsec\ &  19.288 &  19.118 &  19.941 &  0.06 & 2 \\
\enddata
\tablenotetext{a}{1=star; 2=questionable star/cluster classification; 3=detected in only one filter}
\end{deluxetable}

\end{document}